\documentclass[twocolumn,aps,showpacs,prb,tightenlines,amsmath,amssymb,superscriptaddress]{revtex4}
\usepackage{graphicx}
\usepackage{amssymb}
\usepackage{amsmath}
\usepackage{colordvi}
\begin{document}

\title{Reexamination of spin decoherence in semiconductor quantum dots from
  equation-of-motion approach}
\author{J. H. Jiang}
\affiliation{Hefei National Laboratory for Physical Sciences at
  Microscale, University of Science and Technology of China, Hefei,
  Anhui, 230026, China}
\affiliation{Department of Physics,
University of Science and Technology of China, Hefei,
  Anhui, 230026, China}
\altaffiliation{Mailing Address}
\author{Y. Y. Wang}
\affiliation{Department of Physics,
University of Science and Technology of China, Hefei,
  Anhui, 230026, China}
\author{M. W. Wu}
\thanks{Author to whom correspondence should be addressed}
\email{mwwu@ustc.edu.cn}
\affiliation{Hefei National Laboratory for Physical Sciences at
  Microscale, University of Science and Technology of China, Hefei,
  Anhui, 230026, China}
\affiliation{Department of Physics,
University of Science and Technology of China, Hefei,
  Anhui, 230026, China}

\date{\today}
\begin{abstract}

The longitudinal and transversal spin decoherence times, $T_1$ and $T_2$, in
semiconductor quantum dots are investigated from equation-of-motion
approach for different 
 magnetic fields, quantum dot sizes,
and temperatures.  Various mechanisms,
such as the hyperfine interaction with the surrounding nuclei,
the Dresselhaus spin-orbit coupling together with the
electron--bulk-phonon
interaction, the $g$-factor fluctuations, the direct spin-phonon
coupling due to the phonon-induced strain, and the coaction of the
electron--bulk/surface-phonon
interaction together with the hyperfine interaction are
included. The relative contributions from these spin decoherence mechanisms are
compared in detail. In our calculation, the spin-orbit coupling is
included in each mechanism and is shown to have marked
effect in most cases. The equation-of-motion approach is
applied in studying both the spin relaxation time $T_1$ and the spin
dephasing time $T_2$, either in Markovian or in non-Markovian limit.
 When many levels are involved at finite temperature,
we demonstrate how to obtain the  spin relaxation 
time from the Fermi Golden rule in the limit of weak spin-orbit coupling. 
However, at high temperature and/or for large spin-orbit coupling,
one has to use the equation-of-motion approach when many levels are involved.
Moreover, spin dephasing  can be much more efficient than
spin relaxation at high temperature, though the two
only differs by a factor of two at low temperature. 

\end{abstract}

\maketitle

\section{Introduction}

One of the most important issues in the growing field of spintronics
is quantum information processing in quantum dots (QDs) using
electron spin.\cite{Awschalom,Hans,Loss,Hanson,Taylor0} A main obstacle is
that the electron spin is unavoidably coupled to the environment
(such as, the lattice) which leads to considerable spin decoherence
(including longitudinal and transversal spin 
decoherences).\cite{Amasha,Elzerman} Various
mechanisms, such as, the hyperfine interaction with the surrounding
nuclei,\cite{Paget, Meier} the Dresselhaus/Rashba spin-orbit
coupling (SOC)\cite{Dresselhaus, Yu} together with the
electron-phonon interaction, $g$-factor fluctuations,\cite{Roth} the
direct spin-phonon coupling due to the phonon-induced
strain,\cite{Meier} and the coaction of the hyperfine interaction
and the electron-phonon interaction can lead to the spin
decoherence. There are quite a lot of
theoretical works on spin decoherence in  QD. Specifically,
Khaetskii and Nazarov  analyzed the spin-flip transition rate
using a perturbative approach due to the SOC together with the
electron-phonon
interaction, $g$-factor fluctuations, the direct spin-phonon coupling
due to the phonon-induced strain
qualitatively.\cite{Khaetskii,Khaetskii0,Nazarov}
 After that, the longitudinal spin decoherence time  $T_1$ due to the Dresslhaus
and/or the Rashba SOC together with the electron-phonon interaction
were studied quantitatively in Refs.\ \onlinecite{Woods,Sousa,
Cheng,Bulaev,Golovach, Destefani,Jose,Falko2,Stano,Stano2,Westfahl}. Among these works, Cheng
{\em et al.}\cite{Cheng} developed an exact diagonalization method and
showed that due to the strong SOC, the previous perturbation
method\cite{Khaetskii0, Nazarov,Woods} is
inadequate in describing  $T_1$. Furthermore,
they also showed that, the perturbation method previously used  missed
an important second-order energy correction and would yield qualitatively
wrong results if the energy correction is correctly included and only the
lowest few states are kept as those in 
Refs.\ \onlinecite{Khaetskii0, Nazarov,Woods}.
These results were later confirmed by Destefani and Ulloa.\cite{Destefani}
The contribution of the coaction of the hyperfine interaction and the
electron-phonon
interaction to longitudinal spin decoherence was 
calculated in Refs.\ \onlinecite{Erlingsson} and
\onlinecite{Abalmassov}. In contrast to the longitudinal spin decoherence 
time, there are relatively fewer works on the transversal
spin decoherence time, $T_2$, also referred
to as the spin dephasing time
 (while the longitudinal spin decoherence time is referred
to as the spin relaxation time for short). The spin dephasing time due to
the Dresselhaus and/or the Rashba SOC together with the electron-phonon
interaction was studied by Semenov and Kim\cite{Semenov} and
by Golovach {\em  et al.}.\cite{Golovach} The contributions of the hyperfine
interaction and the $g$-factor fluctuation were studied
in Refs.\ \onlinecite{Glazman, Khaetskii2,Schliemann2, Coish, Cakir,
  Deng, Erlingsson2, Shenvi, Sousa2, Merkulov, Pershin,Witzel3,Yao,Witzel2,Deng2}
and in Ref.\ \onlinecite{Kim} respectively.
 However, a quantitative calculation of 
electron spin decoherence induced by the direct spin-phonon
  coupling due to phonon-induced strain in  QDs is
still missing. This 
is one of the issues we are going to present in this paper.
In brief, the spin relaxation/dephasing due to
various mechanisms has been studied
previously in many theoretical works.
However, almost all of these works only focus 
individually on one mechanism.
Khaetskii and Nazarov 
discussed the effects of different mechanisms on the spin
relaxation time. Nevertheless, their results are only qualitative and
there is no comparison 
of the relative importance of the
 different mechanisms.\cite{Khaetskii,Khaetskii0,Nazarov}
Recently, Semenov and Kim discussed various mechanisms contributed to
the spin dephasing,\cite{Semenov3} where they gave a ``phase
diagram'' to indicate the most important spin dephasing mechanism in
Si QD where the SOC is not important.
However, the SOC is very important 
in GaAs QDs.  To fully understand the 
microscopic mechanisms of spin relaxation and
dephasing, and to achieve control over the spin coherence in 
QDs,\cite{Witzel,DeSousa,JHJiang} one
needs to gain insight into the
relative importance of each mechanism
to $T_1$ and $T_2$  under various conditions. This is one of the main
purposes of this paper.

Another issue we are going to address relates to
different approaches used in the study of the spin relaxation time.
The Fermi-Golden-rule approach, which is widely used in the
literature, can be used in calculation of the relaxation time
$\tau_{i\to f}$ between any initial state $|i\rangle$ and final state
$|f\rangle$.\cite{Roth,Khaetskii,Khaetskii0,Glavin,Nazarov,Woods,Cheng,Destefani,lv,Wang,Erlingsson,Abalmassov,Falko2,Sousa,Bulaev,Stano,Stano2} 
However, the problem is that when the process of the spin relaxation relates to
many states, ({\em e.g.},  when temperature is high, the
electron can distribute over many states), one should find 
a proper way to average over the relaxation times ($\tau_{i
  \to f}$) of the involved processes to give the total spin relaxation
time ($T_1$). What makes it difficult in GaAs QDs, is that all the
states are {\em impure} spin states with {\em different} expectation
values of spin. In the existing literature, spin relaxation time is
given by the average of the relaxation times of processes from the
initial state $|i\rangle$ to the final state $|f\rangle$ (with
opposite majority spin of $|i\rangle$)
weighted by the distribution of the initial states $f_{i}$,\cite{Cheng,lv,Wang}
i.e., 
\begin{equation}
\label{goldenT}
T_{1}^{-1}=\sum_{if} f_{i} \tau_{i \to
  f}^{-1}\ .
\end{equation}  
This is a good approximation
in the limit of small SOC as each state only carries a small
amount of minority spin.
 However, when the SOC is very strong which happens at high levels,
it is difficult to find the proper way to perform the
average. We will show that Eq.\ (\ref{goldenT}) is not adequate
any more. Thus, to investigate both $T_1$ and
$T_2$ at finite temperature for arbitrary strength of SOC, we develop
an equation-of-motion approach for the many-level system via
projection operator technique\cite{Argyres} in the Born approximation. With the
rotating wave approximation, we obtain a 
formal solution to the equation of motion. By assuming a proper
initial distribution, we can calculate the evolution of the
  expectation value of spin. We thus obtain the spin
relaxation/dephasing time by the $1/e$ decay of the expectation
 value of spin operator $\langle S_z\rangle$ or $|\langle S_{+}\rangle|$ 
(to its equilibrium 
value), with $S_{+}\equiv S_{x}+iS_{y}$. 
With this approach, we
 are able to study spin relaxation/dephasing for various temperature,
 SOC strength, and magnetic field. 

For quantum information processing based on electron spin in
QDs, the quantum phase coherence is very
  important. Thus, the spin dephasing time is a more relevant
  quantity. There are two kinds of spin dephasing times: the ensemble
  spin dephasing time $T_2^{\ast}$ and the irreversible spin dephasing
  time $T_2$. For a direct measurement of an ensemble of QDs\cite{Braun} 
or an average over many measurements at different
  times where the configurations of the environment have been
  changed,\cite{Koppens,Petta,Koppens2} it gives the ensemble spin
  dephasing time $T_2^{\ast}$. The irreversible spin dephasing 
  time $T_2$ can be obtained by spin echo measurement.\cite{Petta,Koppens2}
A widely discussed source which leads to both $T_2^{\ast}$ and $T_2$
 is the hyperfine interaction between the electron spin and the
 nuclear spins of the lattice. As it has been found that $T_2^{\ast}$
  is around 10\ ns, which is too short and makes a practical quantum
 information processing difficult in electron spin based qubits in
QDs. Thus a spin echo technique is needed to remove the
 free induction decay and to elongate the spin dephasing
  time. Fortunately, this technique has been achieved  first by Petta
  {\em et al.} for two electron triplet-singlet system and then by
Koppens {\em et al.} for a single electron spin 
  system. The achieved spin dephasing time is $\sim 1$\ $\mu$s, which
 is much longer than $T_2^{\ast}$. We therefore discuss only the 
irreversible spin dephasing time $T_2$ throughout the paper, 
i.e., we do not consider the free induction decay
in the hyperfine-interaction-induced spin dephasing.

It is further noticed that Golovach {\em et al.} have shown that
the spin dephasing time $T_2$ is two times  the spin
relaxation time $T_1$.\cite{Golovach} However, as temperature increases,
this relation does not hold. Semenov and Kim on the other hand 
reported that the spin
dephasing time is much smaller than the spin
relaxation time.\cite{Semenov} In this paper, we calculate
the temperature dependence of  the ratio of the spin relaxation
time to the spin dephasing time and analyze
the underlying physics.

This paper is organized as follows: In Sec.\ II, we present our model
and formalism of the equation-of-motion approach.
We also briefly introduce all the
spin decoherence mechanisms considered in our calculations.
In Sec.\ III we present our numerical results to
indicate the contribution of each spin decoherence mechanism
to spin relaxation/dephasing time under
various conditions based on the equation-of-motion approach.
Then we study the problem of how to obtain the spin relaxation time
from the Fermi Golden rule when many levels are involved in Sec.\ IV. 
The temperature dependence of $T_{1}$ and $T_{2}$ is
investigated in Sec.\ V. We conclude in Sec.\ VI.

\section{Model and Formalism}

\subsection{Model and Hamiltonian}

We consider a QD system, where the QD is confined by a
parabolic potential $V_c(x,y) = \frac{1}{2}m^{\ast} \omega_0^2 (x^2+y^2)$ in the
quantum well plane. The width of the quantum well is $a$.
The external magnetic field ${\mathbf B}$ is along $z$
direction, except in Sec.\ IV. The total 
Hamiltonian of the system of electron together with the lattice is:
\begin{equation}
H_T=H_{e}+H_{L}+H_{eL}\ ,
\end{equation}
where $H_{e}$, $H_{L}$, $H_{eL}$ are the Hamiltonians of the electron,
the lattice and their interaction, respectively. The electron
  Hamiltonian is given by
\begin{equation}
H_{e}= \frac{\mathbf{P}^{2}}{2m^\ast} +
V_c({\mathbf r}) + H_Z + H_{SO}
\label{hamiltonian}
\end{equation}
where $\mathbf{P}=-i\hbar\mathbf{\mbox{\boldmath
    $\nabla$\unboldmath}}+\frac{e}{c}\bf{A}$ with 
$\bf{A}=({B_{\perp}}/2)(-y,x)$ ($B_{\perp}$ is the magnetic field
along $z$ direction), $H_Z=\frac{1}{2}g\mu_B
{\mathbf B} \cdot\mbox{\boldmath $\sigma$\unboldmath}$ is the Zeeman
energy with $\mu_B$ the Bohr magneton, and $H_{SO}$ is the Hamiltonian of SOC.
In GaAs, when the quantum well width is small or the gate-voltage
along the growth direction is small, the Rashba SOC is
unimportant.\cite{flat} Therefore, only the
Dresselhaus term\cite{Dresselhaus} contributes to $H_{SO}$. When the
quantum well width is smaller than the QD radius, the
dominant term in the Dresselhaus SOC reads
\begin{equation}
  H_{so}= \frac{\gamma_{0}}{\hbar^{3}}
\langle P_{z}^{2} \rangle_{{\lambda_0}} (-P_{x}\sigma_{x}+P_{y}\sigma_{y})\ ,
\end{equation}
with $\gamma_{0}$ denoting the Dresselhaus coefficient, 
$\lambda_0$ being the
quantum well subband index of the lowest one and $\langle
P_{z}^{2} \rangle_\lambda \equiv -\hbar^2\int \psi_{z\lambda}^\ast(z)\partial^2/\partial
z^2\psi_{z\lambda}(z)dz$. The Hamiltonian of the lattice consists of
two parts $H_{L}=H_{ph} + H_{nuclei}$, where
$H_{ph}=\sum_{{\mathbf q}\eta}\hbar\omega_{{\mathbf q}\eta}
a^{\dagger}_{{\mathbf q}\eta} a_{{\mathbf q}\eta}$
($a^{\dagger}$/$a$ is the phonon creation/annihilation operator)
describes the vibration of the lattice and
$H_{nuclei}=\sum_{j}\gamma_I{\mathbf B}\cdot {\mathbf I}_{j}$
($\gamma_I$ is the gyro-magnetic ratios of the nuclei and
${\mathbf I}_{j}$ is the spin of the $j$-th nucleus) 
describes the precession of the nuclear spins of the lattice in the
external magnetic field. We focus on the spin dynamics due to hyperfine
interaction at a time scale much smaller than the nuclear dipole-dipole
 correlation time ($10^{-4}$\ s in GaAs\cite{Merkulov,Coish}), where the
 nuclear dipole-dipole interaction can be ignored. Under this
  approximation, the equation of motion for the reduced electron system can be
  obtained which only depends on the initial distribution of the
  nuclear spin bath.\cite{Coish} The interaction
between the electron and the 
lattice also has two parts $H_{eL}=H_{eI}+H_{e-ph}$, where
$H_{eI}$ is the hyperfine
interaction between the electron and nuclei and
$H_{e-ph}$ represents the electron-phonon interaction which is further
composed of the electron--bulk-phonon (BP) interaction $H_{ep}$,
the direct spin-phonon coupling due to
the phonon-induced strain $H_{strain}$ and 
 phonon-induced $g$-factor fluctuation $H_g$.

\subsection{Equation-of-motion approach}

The equations of motion can describe both the coherent and the dissipative
dynamics of the electron system. When the quasi-particles of the bath
relax much faster than the electron system, the Markovian
approximation can be made; otherwise the kinetics is the non-Markovian.
For electron-phonon coupling, due to
the fast relaxation of the phonon bath and the weak
electron-phonon scattering, the kinetics of the electron
is Markovian. Nevertheless, as the nuclear spin bath relaxes
much slower than the electron spin,  the kinetics due to the coupling
with nuclei is of non-Markovian
type.\cite{Glazman,Schliemann2,Coish} It is further noted that 
there is also a contribution from
the coaction of the electron-phonon and electron-nuclei couplings, 
which is a fourth order coupling
to the bath. For this contribution,
the decoherence of spin is
mainly controlled by the electron-phonon scattering
while the hyperfine (Overhauser) field\cite{Slichter} acts as a
static magnetic field. Thus, this fourth order coupling is also
Markovian. Finally, since the electron orbit relaxation is much faster
than the electron spin relaxation,\cite{Fujisawa} we  always assume a
thermo-equilibrium initial distribution of the orbital degrees
of freedom.

Generally, the interaction between the electron and the
 quasi-particle of the bath is weak. Therefore the first Born
  approximation is adequate in the treatment of the interaction.
Under this approximation, the equation of
motion for the electron system coupled to the lattice
  environment can be obtained with the help
of the projection operator technique.\cite{Argyres} 
We then assume a sudden
approximation so that the initial distribution of the whole system
is $\rho(t=0)=\rho^e(0)\otimes\rho^{L}(0)$, where $\rho^e$, $\rho^{L}$ is the
density matrix of the system and bath respectively. 
This approximation
corresponds to a sudden injection of the electron into the quantum
dot, which is reasonable for genuine experimental setup.\cite{Coish}
As the initial distribution of the the lattice $\rho^{L}(0)$
  commutates with the Hamiltonian of the lattice $H_{L}$, the
  equation of motion can be written as
\begin{eqnarray}
  &&\hspace{-0.4cm}\frac{d \rho^{e}(t)}{d t}= -\frac{i}{\hbar} [H_{e}+ \mbox{Tr}_{L}(H_{eL}\rho^{L}(0)), \rho^{e}(t)]
  \nonumber\\
  &&\hspace{0.6cm}\mbox{}-\frac{1}{\hbar^2}\int_{0}^{t}d\tau  \mbox{Tr}_{L} [H_{eL},
 U_0(\tau)(\hat{\cal{P}}[H_{eL}, \rho^{e}(t-\tau) \nonumber\\ &&\hspace{2.5cm}\otimes
 \rho^{L}(0)]) U_0^{\dagger}(\tau)]\ ,
\end{eqnarray}
where $\rho^e(t)$ is the density operator of the electron system at
time $t$, $\mbox{Tr}_{L}$ stands for the trace
over the lattice degree of freedom, 
and $U_0(\tau)=e^{-i(H_{L}+H_{e})\tau}$ is time-evolution
operator without $H_{eL}$.
$\hat{\cal{P}}=\hat{1}-\rho^{L}(0)\otimes\mbox{Tr}_{L}$ is the
  projection operator.
The initial distribution of the phonon system is chosen to be the
 thermo-equilibrium distribution.\cite{Golovach} It has been shown by previous
theoretical studies that the initial state of the nuclear spin bath is 
 crucial to the spin dephasing and
  relaxation.\cite{Glazman,Schliemann2,Coish}
Although it may take a
long time ({\em e.g.}, seconds) for the nuclear spin system to relax to its
  thermo-equilibrium state, one can still assume that its initial state 
 is the thermo-equilibrium one. This assumption corresponds to the
 genuine case of enough long waiting time during every individual
 measurement. For a typical setup at above 10\ mK
and with about  10\ T external magnetic field, the thermo-equilibrium
distribution is a distribution with equal probability on every
state. For these initial distributions of phonons and nuclear spins,
the term $\mbox{Tr}_{L}(H_{eL}\rho^{L}(0))$ is zero. Thus,
\begin{equation}
\hat{\cal{P}}[H_{eL}, \rho^{e}(t-\tau) \otimes \rho^{L}(0)]
= [H_{eL}, \rho^{e}(t-\tau) \otimes \rho^{L}(0)] \ .
  \end{equation}
 The equation of motion is then simplified to,
\begin{eqnarray}
 &&\hspace{-0.5cm}\frac{d \rho^{e}(t)}{d t}= -\frac{i}{\hbar} [H_{e}, \rho^{e}(t)]
-\frac{1}{\hbar^2}\int_{0}^{t}d\tau  \mbox{Tr}_{L} [H_{eL},
    [H^{I}_{eL}(-\tau), \nonumber\\ &&
    U^{e}_0(t)
\rho^{Ie}(t-\tau){U^{e}_0}^{\dagger}(t) 
\rho^{L}(0)]]\ ,
 \label{kinetic}
\end{eqnarray}
 where $H^{I}_{eL}$ and $\rho^{Ie}$ are the corresponding operators
  ($H_{eL}$ and $\rho^{e}$ ) in the interaction picture, and
  $U^{e}_0(t)=e^{-iH_{e}t}$ is the time-evolution operator of $H_{e}$.
It should be further noted that the first Born approximation can
not fully account for the non-Markovian dynamics due to the hyperfine
interaction with nuclear spins.\cite{Fulton,Coish} Only when the Zeeman
splitting is much larger than the fluctuating Overhauser shift, the
first Born approximation is adequate. For GaAs QDs, this
requires  $B\gg3.5$\ T.\cite{Coish} In this
 paper, we focus on the study of spin dephasing for the high magnetic
 field regime of $B>3.5$\ T under the first Born approximation, where the 
 second Born approximation only affects
 the long-time behavior.\cite{Coish} Later we will argue that
 this correction of long time dynamics changes the spin dephasing
time very little.

\subsubsection{Markovian kinetics}

The kinetics due to the coupling with phonons can be investigated within
the Markovian approximation, where the equation of motion reduces to, 
\begin{eqnarray}
  &&\hspace{-0.7cm}\frac{d \rho^{e}(t)}{d t}= -\frac{i}{\hbar} [H_{e}, \rho^{e}(t)]
 -\frac{1}{\hbar^2}\int_{0}^{t}d\tau  \mbox{Tr}_{ph} [H_{e-ph},\nonumber\\
 &&\hspace{1.0cm}\mbox{}
    [H^{I}_{e-ph}(-\tau), \rho^{e}(t)\otimes \rho^{ph}(0)]]\ .
\end{eqnarray}
Here $\mbox{Tr}_{ph}$ is the trace over phonon degrees of freedom and
$\rho^{ph}(0)$ is the initial distribution of the phonon bath.
Within the basis of the eigen-states of the electron Hamiltonian,
$\{|\ell\rangle\}$, the above equation  reads,
\begin{eqnarray}
 \hspace{-0.2cm} \frac{d}{d t} \rho_{\ell_1\ell_2}^e &&\hspace{-0.4cm}= -
  i\frac{(\varepsilon_{\ell_1}-\varepsilon_{\ell_2})}{\hbar}
  \rho_{\ell_1\ell_2}^e\nonumber\\
  &&\hspace{-0.05cm} - \Big\{\frac{1}{\hbar^2}
  \int_{0}^{t}\!\!d{\tau} \sum_{\ell_3\ell_4}
\mbox{Tr}_{p}
  (H_{\ell_1\ell_3}^{e-ph}H_{\ell_3\ell_4}^{I\ e-ph}
  \rho_{\ell_4\ell_2}^e\otimes\rho_{eq}^{p} \nonumber\\ &&
  \hspace{0.5cm}-H_{\ell_1\ell_3}^{I\ e-ph}\rho_{\ell_3\ell_4}^e\otimes
  \rho_{eq}^{p}H_{\ell_4\ell_2}^{e-ph}) + H.c.\Big\}\ .
\label{rhoe}
\end{eqnarray}
Here $H_{\ell_1\ell_3}^{e-ph}=\langle
\ell_1|H_{e-ph}|\ell_3\rangle$ and $H_{\ell_1\ell_3}^{I\ e-ph}=\langle
\ell_1|H^{I}_{e-ph}(-\tau)|\ell_3\rangle$.
A general form of the electron-phonon interaction reads
\begin{equation}
H_{e-ph}=\sum_{{\bf q}\eta}\Phi_{{\bf q}\eta}(a_{{\bf
    q}\eta}+a^{\dagger}_{-{\bf q}\eta}) X_{{\bf q}\eta}({\bf
  r},{\mbox{\boldmath $\sigma$\unboldmath}})\ .
\label{ge-ph}
\end{equation}
Here, $\eta$ represents the phonon branch index; $\Phi_{{\bf q}\eta}$ is the
matrix element of the electron-phonon interaction; $a_{{\bf q}\eta}$
is the phonon annihilation operator; $X_{{\bf q}\eta}({\bf
  r},{\mbox{\boldmath $\sigma$\unboldmath}})$ denotes a function of
electron position and spin.
Substituting this into Eq.\ (\ref{rhoe}), we obtain, after
 integration within the
Markovian approximation,\cite{JHJiang}
\begin{eqnarray}
  \frac{d}{d t} \rho_{\ell_1\ell_2}^e &=&
  i\frac{(\varepsilon_{\ell_1}-\varepsilon_{\ell_2})}{\hbar}
  \rho_{\ell_1\ell_2}^e\nonumber\\ &&
  - \Big\{\frac{\pi}{\hbar^2} \sum_{\ell_3\ell_4}\sum_{{\bf q}\eta}
   |\Phi_{{\bf q}\eta}|^{2}\{ X_{\ell_1\ell_3}^{{\bf
      q}\eta}X_{\ell_4\ell_3}^{{\bf
      q}\eta\ast}\rho_{\ell_4\ell_2}^e \nonumber\\
  && \hspace{0.5cm} \times C_{{\bf q}\eta}
  (\varepsilon_{\ell_4}-\varepsilon_{\ell_3})\mbox{} -X_{\ell_4\ell_2}^{{\bf
      q}\eta}X_{\ell_3\ell_1}^{{\bf
      q}\eta\ast}\rho_{\ell_3\ell_4}^e \nonumber\\
  && \hspace{0.5cm} \times C_{{\bf
      q}\eta}(\varepsilon_{\ell_3}-\varepsilon_{\ell_1})\} + H.c.\
  \Big\}
\label{em}
\end{eqnarray}
in which $X_{\ell_1\ell_2}^{{\bf q}\eta} = \langle\ell_1|X_{{\bf
    q}\eta}({\bf r},{\mbox{\boldmath $\sigma$\unboldmath}})|\ell_2\rangle$,
and $C_{{\bf
  q}\eta}(\Delta\varepsilon) = \bar{n}(\omega_{{\bf
 q}\eta})\delta(\Delta\varepsilon+\omega_{{\bf q}\eta})+[\bar{n}(\omega_{{\bf
    q}\eta})+1]\delta(\Delta\varepsilon-\omega_{{\bf q}\eta})$. Here
$\bar{n}(\omega_{{\bf q}\eta})$ represents the Bose distribution function.
Equation\ (\ref{em})  can be written in a more compact form
\begin{equation}
  \frac{d}{d t} \rho_{\ell_1\ell_2}^e =
  - \sum_{\ell_3\ell_4}
  \Lambda_{\ell_1\ell_2\ell_3\ell_4}
  \rho_{\ell_3\ell_4}^e\ ,
\end{equation}
which is a linear differential equation. This equation can be solved
by diagonalizing ${\bf \Lambda}$. Given an initial distribution
$\rho_{\ell_1\ell_2}^e(0)$, the density matrix
$\rho_{\ell_1\ell_2}^e(t)$ and the expectation value of any physical
quantity $\langle O\rangle_t = \mbox{Tr}({\hat O}{\bf {\rho}}^e(t))$ at time
$t$ can be obtained:\cite{JHJiang}
\begin{eqnarray}
\langle O\rangle_t &=& \mbox{Tr}({\hat O}{\bf {\rho}}^e) \nonumber\\
  &=&\sum_{\ell_1\cdots\ell_6}
  \langle\ell_2|{\hat O}|\ell_1\rangle
  {\bf P}_{(\ell_1\ell_2)(\ell_3\ell_4)}\nonumber\\
&&\mbox{}\times e^{-{\bf
 \Gamma}_{(\ell_3\ell_4)}t} {\bf
  P}_{(\ell_3\ell_4)(\ell_5\ell_6)}^{-1}
\rho_{\ell_5\ell_6}^e(0)
\label{Oeq}
\end{eqnarray}
with ${\bf \Gamma}={\bf P}^{-1}{\bf \Lambda}{\bf P}$ being the diagonal
matrix and ${\bf P}$ representing the transformation matrix. To study spin
dynamics, we calculate $\langle
  S_z \rangle_t$ ($|\langle S_{+} \rangle_t|$) and define the spin
relaxation (dephasing) time as the time when  
$\langle S_z \rangle_t$ ($|\langle S_{+}
  \rangle_t|$) decays to $1/e$ of its initial value (to its equilibrium 
value).

\subsubsection{Non-Markovian kinetics}
Experiments have already shown that for a large ensemble of
  quantum dots or for an ensemble of many measurements on the same
  quantum dot at different times, the spin dephasing time due to
  hyperfine interaction is quite short, 
$\sim 10$\ ns.\cite{Braun,Koppens,Petta,Koppens2} 
This rapid spin dephasing is caused by the 
  ensemble broadening of the precession frequency due to the hyperfine
  fields.\cite{Merkulov} When the external magnetic field is much
  larger than the random Overhauser field, the rotation due to the
  Overhauser field perpendicular to the magnetic field is blocked. Only
  the broadening of the Overhauser field parallel to the magnetic field
  contribute to the spin dephasing. To describe this free induction
  decay for this high magnetic field case, we
write the hyperfine interaction into two parts: $H_{eI}={\mathbf
    h} \cdot {\mathbf S}=H_{eI1}+H_{eI2}$. Here ${\mathbf
    h}=(h_x,h_y,h_z)$ and ${\mathbf S}=(S_x,S_y,S_z)$ are the
  Overhauser field and the electron spin respectively. $H_{eI1}=h_zS_z$ and
  $H_{eI2}= \frac{1}{2}(h_{+}S_{-}+h_{-}S_{+})$ with $h_{\pm}=h_{x}\pm
  ih_{y}$. The longitudinal part $H_{eI1}$ is responsible for the free induction decay,
  while the transversal part $H_{eI2}$ is responsible for high order irreversible
  decay. As the rapid free induction decay can be removed by spin
  echo,\cite{Petta,Koppens2} elongating the spin dephasing time
    to $\sim 1$\ $\mu$s which is more favorable for quantum computation
    and quantum information processing, we then discuss only the
  irreversible decay.
We first classify the states of the nuclear spin system with its
polarization. Then we reconstruct the states within the same class to
make it spatially uniform. These uniformly polarized pure states,
$|n\rangle$'s, are eigen-states of $h_z$. They also form a
complete-orthogonal basis of the nuclear spin system. A formal
expression of $|n\rangle$ is\cite{Coish}
\begin{equation}
|n\rangle = \sum_{m_1\cdots m_N} \alpha^{n}_{m_1\cdots m_N}
 \bigotimes_{j=1}^{N} |I,m_{j} \rangle\ .
\end{equation}
Here $|I,m_{j} \rangle$ denotes the eigen-state of the $z$-component of
the $j$-th nuclear spin $I_{j z}$ with the eigenvalue $\hbar m_{j}$. 
$N$ denotes the number of the nuclei.
The equation of motion for the case with initial nuclear spin state 
$\rho^{ns}_{1}(0)=|n\rangle\langle n|$ is given by\cite{Coish} 
\begin{eqnarray}
\frac{d \rho^{e}(t)}{d t}&=& -\frac{i}{\hbar} [H_{e} + \mbox{Tr}_{ns}(H_{eI}\rho^{ns}_{1}(0)), \rho^{e}(t)]
    \nonumber\\
    &&\hspace{0.0cm}\mbox{}
-\frac{1}{\hbar^2} \int_{0}^{t}d\tau  \mbox{Tr}_{ns} [H_{eI2},
    U^{eI}_0(\tau)\nonumber\\
&&\times[H_{eI2}, \rho^{e}(t-\tau)
\otimes \rho^{ns}_{1}(0)]{U^{eI}_0}^{\dagger}(\tau)]\ .
\label{hyp_eom}
\end{eqnarray}
As in traditional projection operator technique, the dynamics of the
nuclear spin subsystem is incorporated self-consistently in the last
term.\cite{Coish,Argyres} 
Here $\mbox{Tr}_{ns}$ is the trace over nuclear spin degrees of freedom.
$U^{eI}_0(\tau)=\exp[-i\tau(H_e+H_I+H_{eI1})]$. 
The Overhauser field is given by
${\mathbf h}=\sum_{j} A v_{0} {\mathbf I}_{j} \delta({\mathbf r}-
{\mathbf R}_{j})$, where the 
constants $A$ and $v_{0}$ are given later. ${\mathbf I}_{j}$ and
${\mathbf R}_{j}$ are the spin and position of $j$-th nucleus
respectively. As mentioned above, the initial state of the nuclear
spin bath is chosen to be a state with equal probability of each
state, therefore $\rho^{ns}(0) = \sum_{n}1/N_w |n\rangle\langle n|$,
with $N_w=\sum_n 1$ being the number of states of the basis
$\{|n\rangle\}$. To quantify the irreversible decay, we calculate
the time evolution of $S_{+}^{(n)}$ for every case with 
initial nuclear spin state $|n\rangle$.
We then sum over $n$ and obtain 
\begin{equation} 
||\langle S_{+}\rangle_t||=\sum_{n} |\langle S_{+}^{(n)}\rangle_t| .
\end{equation}
It is noted that the summation is performed 
after the absolute value of $\langle S_{+}^{(n)}\rangle_t$.
Therefore, the destructive interference due to the difference in
precession frequency $\omega_{zn}$, which originates from the
longitudinal part of the hyperfine interaction ($H_{eI1}$), is {\em removed}.
We thus use $1/e$ decay of the envelope of $||\langle
S_{+}\rangle_t||$ to describe the irreversible spin dephasing time
$T_2$. Similar description has been used in the irreversible spin dephasing in
semiconductor quantum wells\cite{Wu} and the irreversible inter-band optical
dephasing in semiconductors.\cite{Fossi,Shah}

Expanding Eq.\ (\ref{hyp_eom}) in the basis of $\{|n\rangle\}$, one
obtains, 
\begin{eqnarray}
  \hspace{0.0cm} \frac{d}{d t} \rho_{\ell_1\ell_2}^e &&\hspace{-0.4cm}= -
 \frac{i}{\hbar}\sum_{\ell_3}\Big\{(\varepsilon_{\ell_1}\delta_{\ell_1\ell_3}+H_{n\ell_1;n\ell_3}^{eI1})\rho^e_{\ell_3\ell_2}\nonumber\\
&&\hspace{0.6cm}-\rho^e_{\ell_1\ell_3}(\varepsilon_{\ell_3}\delta_{\ell_3\ell_2}
+H_{n\ell_3;n\ell_2}^{eI1})\Big\} \nonumber\\
  &&\hspace{-1.2cm} - \Big\{\frac{1}{\hbar^2}
  \int_{0}^{t}\!\!d{\tau} \sum_{n_1}\sum_{\ell_3\ell_4}
[H_{n\ell_1;n_1\ell_3}^{eI2}H_{n_1\ell_3;n\ell_4}^{I\ eI2}
\rho_{\ell_4\ell_2}^e(t-\tau)\nonumber\\
&&\hspace{-0.8cm} -H_{n\ell_1;n_1\ell_3}^{I\ eI2}\rho_{\ell_3\ell_4}^e(t-\tau) 
  H_{n_1\ell_4;n\ell_2}^{eI2}]
+ H.c.\Big\}\ .
\end{eqnarray}
Here $H_{n\ell_1;n_1\ell_3}^{eI2}=\langle
n\ell_1|H_{eI2}|n_1\ell_3\rangle$ and $H_{n\ell_1;n_1\ell_3}^{I\ eI2}=\langle
n\ell_1|H^{I}_{eI2}(-\tau)|n_1\ell_3\rangle$.
For simplicity, we neglect the terms concerning different orbital
wavefunctions which are much smaller. For small spin
mixing, assuming an equilibrium distribution in orbital degree of
freedom, under rotating wave approximation, and trace over the
orbital degree of freedom, we finally arrive at 
\begin{eqnarray}
\frac{d}{dt} \langle S_{+}^{(n)}\rangle_t 
&=& i \omega_{zn} \langle S_{+}^{(n)}\rangle_t
- \frac{1}{\hbar^2}\int^{t}_{0} d \tau \{\frac{1}{4}
\sum_{kn^{\prime}} f_{k} (
[h_{+}]_{k{n}{n^{\prime}}} \nonumber\\
&&\mbox{} \times [h_{-}]_{{kn^{\prime}}{n}}+
[h_{-}]_{k{n}{n^{\prime}}} [h_{+}]_{{kn^{\prime}}{n}} )
\nonumber\\ &&\mbox{}
\times\exp[i\tau(\omega_{kn}-\omega_{kn^{\prime}})] \} \langle
S_{+}^{(n)}\rangle_{t-\tau}\ .
\label{hyp_eom2}
\end{eqnarray}
Here $\omega_{zn} = \sum_{k} f_k (E_{zk}/\hbar + \omega_{kn})$ with
$E_{zk}$ representing the electron Zeeman splitting of the $k$-th orbital
level. $[h_i]_{knn^{\prime}}=\langle
n|\langle k|h_{i}|k\rangle| n^{\prime}\rangle$ ($i=\pm,z$).
$\omega_{kn}=[h_z]_{knn} + \epsilon_{nz}$ with $\epsilon_{nz}$ denoting
the nuclear Zeeman splitting which is very small and can be neglected.
By solving the above equation, we obtain  $|\langle
S_{+}^{(n)}\rangle_t|$ for a given $|n\rangle$. We then sum over $n$ and
determine the irreversible spin dephasing time $T_2$ as
 $1/e$ decay of  the envelop of $||\langle S_{+}\rangle_t||$. By noting that only the 
polarization of nuclear spin state $|n\rangle$ determines the
evolution of $|\langle S_{+}^{(n)}\rangle_t|$, the summation over $n$ is then
reduced to summation over polarization which becomes a integration for
large $N$. This integration can be handled numerically.

In the limiting case of zero SOC and very low
temperature, only the lowest two Zeeman sublevels are concerned. The
equation for $\langle S_{+}\rangle_t$ with initial nuclear spin state 
$\rho^{ns}_{1}(0)=|n\rangle\langle n|$ reduce to
\begin{eqnarray} 
&&\hspace{-0.5cm}\frac{d}{dt} \langle S_{+}\rangle_t = i \omega_{zn} \langle S_{+}\rangle_t
- \frac{1}{\hbar^2}\int^{t}_{0} d \tau \{\frac{1}{4} \sum_{n^{\prime}} (
[h_{+}]_{{n}{n^{\prime}}} \nonumber\\
&&\mbox{}\hspace{-0.5cm} \times [h_{-}]_{{n^{\prime}}{n}}+
[h_{-}]_{{n}{n^{\prime}}} [h_{+}]_{{n^{\prime}}{n}} )
\exp[i\tau(\omega_{n}-\omega_{n^{\prime}})] \} \langle S_{+}\rangle_{t-\tau} \nonumber\\
&=&i \omega_{z} \langle S_{+}\rangle_t - \int^{t}_{0} d \tau
\Sigma(\tau) \langle S_{+}\rangle_{t-\tau}\ .
\label{evolution}
\end{eqnarray}
In this equation $\omega_{zn}= (g\mu_{B} B + [h_z]_{nn^{\prime}})/\hbar$,
$[h_\xi]_{nn^{\prime}}=\langle n|\langle 
\psi_1|h_{\xi}|\psi_1\rangle| n^{\prime}\rangle$ ($\xi=\pm,z$ and $\psi_1$ is
the orbital quantum number of the ground state), 
and $\omega_{n}=[h_z]_{nn}$. Similar equation has 
been obtained by Coish and Loss,\cite{Coish}  
and later by Deng and Hu\cite{Deng} at very low temperature such 
that only the lowest two Zeeman sublevels are considered.
 Coish and Loss also presented an efficient way
to evaluate $\Sigma(\tau)$ in terms of
their Laplace transformations, $\Sigma(s)= \int_{0}^{\infty} d\tau e^{-s\tau}
\Sigma(\tau)$. They gave,
\begin{eqnarray}
\Sigma (s) &=&\frac{1}{4\hbar^2} \sum_{n^{\prime}} (
[h_{+}]_{{n}{n^{\prime}}} [h_{-}]_{{n^{\prime}}{n}}\nonumber\\
&&\mbox{} +
[h_{-}]_{{n}{n^{\prime}}} [h_{+}]_{{n^{\prime}}{n}} )
/(s - i \delta \omega_{nn^{\prime}})\ ,
\label{selfe}
\end{eqnarray}
with $\delta \omega_{nn^{\prime}} = \frac{1}{2}
(\omega_{n} - \omega_{n^{\prime}})$. With the help of this technique, we are
  able to investigate the spin dephasing due to
the hyperfine interaction.

\subsection{Spin decoherence mechanisms}

In this subsection we briefly summarize all the spin decoherence
mechanisms. It is noted that the SOC modifies all the mechanisms.
This is because the SOC modifies the Zeeman splitting\cite{Cheng}
 and the spin-resolved
eigen-states of the electron Hamiltonian, it hence greatly changes
the effect of the electron-BP scattering.\cite{Cheng}
These two modifications, especially the modification of the Zeeman
splitting, also change the effect of other mechanisms, such as, the
direct spin-phonon coupling due to the phonon-induced strain, the
$g$-factor fluctuation, the coaction of the electron-phonon
interaction and the hyperfine interaction. In the literature, except
for the electron-BP scattering, the effects from the SOC
are neglected except for the work by Woods {\em et
al}.\cite{Woods} in which the spin relaxation time
between the two Zeeman sub-levels of the lowest electronic state due to
the phonon-induced strain is investigated. However, the perturbation
method they used does not include the important second-order energy
correction.  In our investigation, the effects of the SOC are 
included in all the mechanisms
and we will show that they  lead to marked effects in
most cases.

\subsubsection{SOC together with electron-phonon scattering}

As the SOC mixes different spins, the electron-BP scattering
can induce spin relaxation and dephasing.
The electron-BP coupling is given by
\begin{equation}
  H_{ep}=\sum_{{\mathbf q}\eta} M_{{\mathbf q}\eta} (a_{{\mathbf q}\eta} + a^{\dagger}_{-{\mathbf q}
    \eta})e^{i {\mathbf
      q} \cdot {\mathbf r}}\ ,
 \label{ep}
\end{equation}
where $M_{{\mathbf q}\eta}$ is the matrix element of the electron-phonon
interaction. In the general form of the electron phonon interaction $H_{e-ph}$,
$\Phi_{{\bf q}\eta}=M_{{\mathbf q}\eta}$ and $X_{{\bf q}\eta}({\bf
  r},{\mbox{\boldmath $\sigma$\unboldmath}})=e^{i {\mathbf
      q} \cdot {\mathbf r}}$.
$|M_{{\mathbf q}sl}|^{2} = \hbar \Xi^{2}q/2 \rho v_{sl}V$ for the
electron-BP
coupling due to the deformation potential. For the
piezoelectric coupling,
$|M_{{\mathbf q}  pl}|^{2}= (32\hbar \pi^{2}
e^{2}e_{14}^{2}/\kappa^{2} \rho v_{sl}V)[(3q_{x}q_{y}q_{z})^{2}/q^{7}]$ for the longitudinal phonon mode
and   $\sum_{j=1,2}|M_{{\mathbf q}pt_j}|^{2} =[32\hbar \pi^{2}
e^{2}e_{14}^{2}/(\kappa^{2} \rho
v_{st}q^{5}V)][q_{x}^{2}q_{y}^{2}+q_{y}^{2}q_{z}^{2} +
q_{z}^{2}q_{x}^{2} -(3q_{x}q_{y}q_{z})^{2}/q^{2}]$ for the two transverse modes.
Here $\Xi$
stands for the acoustic deformation potential;
$\rho$ is the GaAs volume density; $V$ is the volume of the lattice; $e_{14}$
is the piezoelectric constant and $\kappa$ denotes
the static dielectric constant. The acoustic phonon spectra
$\omega_{{\mathbf q}ql} = v_{sl} q$ for the longitudinal mode and
$\omega_{{\mathbf q}pt} = v_{st}q$ for the transverse mode with
$v_{sl}$ and $v_{st}$ representing the corresponding sound velocities.

Besides the electron-BP scattering, electron also
couples to
vibrations of the confining potential, {\em i.e.}, the
surface-phonons,\cite{Abalmassov}
\begin{equation}
\delta V ({\mathbf r}) = - \sum_{{\bf q}\eta} \sqrt{\frac{\hbar}{2 \rho \omega_{{\bf q} \eta} V}}
  (a_{{\bf q} \eta}+ a_{- {\bf q} \eta}^{\dagger})
\mbox{\boldmath$\epsilon$\unboldmath}_{{\bf q}\eta} \cdot {\mathbf \nabla}_{{\bf r}} V_c({\bf
    r})\ ,
\end{equation}
in which \boldmath$\epsilon$\unboldmath$_{{\bf q}\eta}$ is the polarization vector
 of a phonon mode with wave-vector ${\mathbf q}$ in branch $\eta$.
 However, this contribution is much smaller
than the electron-BP coupling. Compared to the coupling due to the deformation
potential for example, the ratio of the two coupling strengths is
$\approx \hbar \omega_{0} / \Xi q l_0 $\ , where $l_0$ is the
characteristic length of the quantum dot and $\hbar \omega_{0}$ is the
orbital level splitting. The phonon wave-vector $q$ is determined by
the energy difference between the final and  initial states of the
transition. Typically  phonon transitions between Zeeman
sublevels and different orbital levels, $q l_0$ ranges from $0.1$
to $10$. Bearing in mind that $\hbar \omega_{0}$ is
about 1\ meV while  $\Xi=7$\ eV in GaAs, $ \hbar \omega_{0}/ \Xi q l_0$
is about $10^{-3}$. The piezoelectric coupling is of the same order
 as the deformation potential. Therefore spin decoherence due to the
electron--surface-phonon coupling is negligible.

\subsubsection{Direct spin-phonon coupling due to phonon-induced strain}

The direct spin-phonon coupling due to the phonon-induced strain is given
by\cite{Dyakonov}
\begin{equation}
H_{strain}= \frac{1}{2}{\mathbf h}^{s} (\mathbf p) \cdot
{\mbox{\boldmath $\sigma$\unboldmath}}\ ,
\label{strain1}
\end{equation}
where $h_{x}^{s}= -D p_{x} ({\epsilon}_{yy}
-{\epsilon}_{zz})$,
$h_{y}^{s}= -D p_{y} ({\epsilon}_{zz} -{\epsilon}_{xx})$ and
$h_{z}^{s}= -D p_{z} ({\epsilon}_{xx} -{\epsilon}_{yy})$\ with
$\mathbf p = (p_x,p_y,p_z)=-i\hbar{\mbox{\boldmath $\nabla$\unboldmath}}$\, and
$D$ being the material strain constant. ${\epsilon}_{ij}$
($i,j = x,y,z$) can be expressed by the phonon creation and annihilation
operators:
\begin{eqnarray}
{\epsilon}_{ij} = \sum_{{\mathbf q}\eta=l,t_1,t_2} \frac{i}{2}
\sqrt{\frac{\hbar}{ 2 \rho {\omega_{{\mathbf q} \eta}V}}  }
(a_{{\mathbf q} ,\eta} + a_{-{\mathbf q}, \eta}^{+}) (\xi_{i
  \eta} q_{j} \nonumber \\
&&\hspace{-2.8cm}\mbox{} + \xi_{j\eta} q_{i}) e^{i {\mathbf
    q}\cdot {\mathbf r}}\ ,
\label{strain}
\end{eqnarray}
in which $\xi_{il}= q_{i}/q$ for the longitudinal phonon mode and
$(\xi_{x t_{1}},\xi_{y t_{1}}, \xi_{z t_{1}}) = (q_{x}q_{z},
q_{y}q_{z}, - q_{\|}^{2})/ q q_{\|}$, $(\xi_{x t_{2}},
\xi_{y t_{2}},\xi_{z t_{2}}) = (q_{y}, -q_{x}, 0)/q_{\|}$ for the two transverse phonon
modes with  $q_{\|} = \sqrt{q_{x}^{2}+q_{y}^{2}}$. Therefore,
in the general form of electron-phonon interaction $H_{e-ph}$, $\Phi_{{\mathbf
  q}\eta}= -i D \sqrt{\hbar/(32\rho  \omega_{{\mathbf
  q} \eta} V)}$ and $X_{{\mathbf q}\eta}({\mathbf
  r},{\mbox{\boldmath $\sigma$\unboldmath}})=\sum_{ijk}
\epsilon_{ijk}(\xi_{j\eta} q_{j} - \xi_{k\eta}q_{k} )p_{i} e^{i {\mathbf
    q} \cdot {\mathbf r}}\sigma_{i}$ with
$\epsilon_{ijk}$ denoting the Levi-Civita tensor.

\subsubsection{$g$-factor fluctuation}

The spin-lattice interaction via phonon modulation of the $g$-factor
is given by\cite{Roth}
\begin{equation}
H_{g}= \frac{\hbar}{2}\sum_{ijkl=x,y,z} A_{ijkl} \mu_{B} B_{i}
\sigma_{j} {\epsilon}_{kl}\ ,
\end{equation}
where ${\epsilon}_{kl}$ is given in Eq.\ (\ref{strain})
and $A_{ijkl}$ is a tensor determined by the material.
Therefore in $H_{e-ph}$, $\Phi_{{\bf q}\eta}=i \sqrt{\hbar/(32\rho
  \omega_{{\mathbf
  q} \eta} V)}$ and $X_{{\bf q}\eta}({\mathbf
  r},{\mbox{\boldmath $\sigma$\unboldmath}}) = \sum_{i,j,k,l}
A_{i,j,k,l} \mu_{B} B_i (\xi_{k\eta} q_{k} - \xi_{l\eta}q_{l} )
\sigma_{j} e^{i {\mathbf q} \cdot {\mathbf r}}$.
Due to the axial symmetry with respect to the $z$-axis, and keeping in
mind that the external magnetic field is along the $z$ direction, the
only finite element of $H_{g}$ is $H_{g}=[(A_{33}-A_{31}){\epsilon}_{zz} +
A_{31}\sum_{i}{\epsilon}_{ii}]\hbar{\mu_{B} B}\sigma_{z}/2$ with $A_{33}=A_{zzzz}$,
$A_{31}=A_{zzxx}$ and $A_{66}= A_{xyxy}$. $A_{33} + 2A_{31}
  =0$.\cite{Kim}

\subsubsection{Hyperfine interaction}

The hyperfine interaction between the electron and nuclear spins is\cite{Abragam} 
\begin{equation}
H_{eI}({\mathbf r})=\sum_{j} A v_{0} {\mathbf S} \cdot {\mathbf
  I}_{j} \delta({\mathbf r}- {\mathbf R}_{j})\ ,
\label{hyperfine}
\end{equation}
where ${\bf S}=\hbar \mbox{\boldmath
    $\sigma$\unboldmath}/2$ and ${\bf I}_j$ are the electron
and nucleus spins respectively, $v_{0}=a_0^3$ is the
volume of the unit cell with $a_{0}$
representing the crystal lattice parameter, and ${\mathbf r}$ (${\mathbf R}_{j}$) denotes
the position of the electron (the $j$-th nucleus). $A=4 \mu_{0}\mu_{B}
\mu_{I} /(3 I v_{0})$ is the hyperfine coupling constant with
$\mu_{0}$, $\mu_B$ and $\mu_I$  representing the permeability of
    vacuum, the Bohr
magneton and the nuclear magneton separately.

As the Zeeman splitting of the electron is much larger (three orders
of magnitude
larger) than that of the nucleus spin, to conserve the energy for the
spin relaxation processes, there must be phonon-assisted transitions
when considering the spin-flip processes. Taking
into account directly the BP induced motion of nuclei spin of
the lattice leads to a new spin relaxation mechanism:\cite{Abalmassov}
\begin{equation}
V_{eI-ph}^{(1)} ({\mathbf r}) = - \sum_{j} A v_{0} {\mathbf S} \cdot {\mathbf I}_{j}
({\mathbf u} ({\mathbf R}_{j}^{0}) \cdot {\mathbf \nabla}_{{\mathbf r}}) \delta ({\mathbf r} -
{\mathbf R}_{j})\ ,
\label{1st-order}
\end{equation}
where ${\mathbf u}({\mathbf R}_{j}^{0})= \sum_{{\mathbf q} \eta}
  \sqrt{ \hbar/ (2 \rho \omega_{{\mathbf q} \eta} {v_{0}})} (a_{{\mathbf q}
  \eta} + a_{{\mathbf q} \eta}^{\dagger}) \epsilon_{{\mathbf q} \eta}
  e^{i {\mathbf q} \cdot {\mathbf R}_{j}^{0}}$ is the lattice
  displacement vector.
Therefore using the notation of Eq.\ (\ref{ge-ph}),
$\Phi=\sqrt{\hbar/(2\rho V \omega_{{\mathbf q} \eta})}$
  and $X_{{\mathbf q} \eta} = \sum_{j} A v_{0} {\mathbf S} \cdot
  {\mathbf I}_{j} {\mathbf \nabla}_{{\mathbf r}} \delta ({\mathbf r}-
  {\mathbf R}_{j})$.
The second-order process of the
surface phonon and the BP together with the hyperfine
  interaction also leads to spin relaxation:
\begin{eqnarray}
V_{eI-ph}^{(2)} ({\mathbf r}) &=& 
|\ell_{2}\rangle\Big\{ \sum_{m\not=\ell_1} \frac{ \langle
\ell_{2} | \delta V_c({\mathbf r}) |m \rangle \langle m |
H_{eI}({\mathbf r}) |\ell_{1}
\rangle }{ \varepsilon_{\ell_{1}} -\varepsilon_{m}} \nonumber \\
&&\mbox{} \hspace{-1.0cm}+ \sum_{m\not=\ell_2} \frac{ \langle
\ell_{2} | {H_{eI} ({\mathbf r})} |m \rangle \langle m | \delta V_c({\mathbf r}) |\ell_{1}
\rangle }{\varepsilon_{\ell_{2}} - \varepsilon_{m}} \Big\} \langle \ell_{1}|\ ,
\end{eqnarray}
and
\begin{eqnarray}
V_{eI-ph}^{(3)} &=&   | \ell_{2} 
\rangle\Big\{ \sum_{m\not=\ell_1}\frac{ \langle
\ell_{2} | H_{ep} |m \rangle \langle m | {H_{eI}({\mathbf r})} |\ell_{1}
\rangle }{ \varepsilon_{\ell_{1}} - \varepsilon_{m}} \nonumber \\
&&\mbox{} \hspace{-0.5cm} +\sum_{m\not=\ell_2} \frac{ \langle
\ell_{2} |{H_{eI} ({\mathbf r})} |m \rangle \langle m |H_{ep}|\ell_{1}
\rangle }{\varepsilon_{\ell_{2}} -\varepsilon_{m}} \Big\} \langle \ell_{1}|\ ,
\label{two-order}
\end{eqnarray}
in which $|\ell_{1}\rangle$ and $|\ell_{2}\rangle$ are the eigen states of
  $H_{e}$.
By using the notations in $H_{e-ph}$, 
$\Phi_{{\mathbf q} \eta} = \frac{i}{\hbar}
  \sqrt{\hbar/(2\rho \omega_{{\mathbf q}\eta} {v_{0}})}$ and
\begin{eqnarray}
&&\hspace{-0.4cm}X_{{\mathbf q} \eta} =|\ell_{2}\rangle \mbox{\boldmath$\epsilon$\unboldmath}_{{\mathbf q} \eta}
\cdot 
  \Big\{\sum_{m\not=\ell_1} \frac{1}{\varepsilon_{\ell_{1}}- \varepsilon_{m}}\langle
\ell_{2} |[H_{e},{\mathbf P}] | m \rangle \sum_{j} A v_{0}
  \nonumber \\
&&\hspace{-0.2cm}\mbox{}
\times \langle m| {\mathbf S} \cdot {\mathbf I}_{j} \delta({\mathbf r} -
 {\mathbf R}_{j}) |\ell_{1} \rangle  + \sum_{m\not=\ell_2} \frac{1}{\varepsilon_{\ell_{2}}-
\varepsilon_{m}} \langle
  m|[H_{e},{\mathbf P}] | \ell_{1} \rangle \nonumber \\
&&\hspace{0.2cm}\mbox{}
\times \sum_{j} A v_{0}
 \langle \ell_{2}| {\mathbf S} \cdot {\mathbf I}_{j} \delta({\mathbf r} -
 {\mathbf R}_{j}) |m \rangle\Big \} \langle \ell_{1}|
\end{eqnarray}
for $V_{eI-ph}^{(2)}$. Similarly  $\Phi_{{\mathbf q} \eta}= M_{{\mathbf
    q} \eta}$ and
\begin{eqnarray}
X_{{\mathbf q} \eta} &=& |\ell_{2} \rangle \Big\{\sum_{m\not=\ell_1}
\frac{\langle \ell_{2}| e^{i {\mathbf q} \cdot {\mathbf r}} | m\rangle}
{\varepsilon_{\ell_{1} }-\varepsilon_{m}}
 \sum_{j} A
v_{0} \langle m| {\mathbf S}\cdot {\mathbf I}_{j} \nonumber \\
&&\mbox{} \times \delta ({\mathbf r}-
{\mathbf R}_{j})|\ell_{1} \rangle + \sum_{m\not=\ell_2}\frac{1}{\varepsilon_{\ell_{2}}-\varepsilon_{m}}
\langle m| e^{i {\mathbf q} \cdot {\mathbf r}} | \ell_{1} \rangle \nonumber \\
&&\mbox{}
\times \sum_j A {v_0}\langle \ell_{2}| {\mathbf S} \cdot {\mathbf I}_{j} \delta ({\mathbf r}-
{\mathbf R}_{j})|m\rangle\Big\} \langle \ell_{1}|
\end{eqnarray}
for $V_{eI-ph}^{(3)}$. Again as the contribution from the surface phonon
is much smaller than that of the BP, $V_{eI-ph}^{(2)}$ can be neglected.
It is noted that, the direct spin-phonon coupling due to the
phonon-induced strain
together with the hyperfine interaction
gives a fourth-order scattering and hence induces a  spin
relaxation/dephasing.
The interaction is
\begin{eqnarray}
V_{eI-ph}^{(4)} &=&   | \ell_{2} \rangle \Big\{ {\sum_{m\not=\ell_1}}\frac{ \langle
\ell_{2} | H_{strain}^{z} |m \rangle \langle m | H_{eI}({\mathbf r}) |\ell_{1}
\rangle }{\varepsilon_{\ell_{1}} -\epsilon_{m}} \nonumber \\
&&\mbox{}\hspace{-0.9cm} + {\sum_{m\not=\ell_2}}\frac{ \langle
\ell_{2} | H_{eI} ({\mathbf r}) |m \rangle \langle m
|H_{strain}^{z}| \ell_{1
\rangle }}{\epsilon_{\ell_{2} } - \epsilon_{m}} \Big\} \langle \ell_{1}|\ ,
\end{eqnarray}
with $H_{strain}^{z}= h_{s}^{z} \sigma_{z} /2$
only changing the electron energy but conserving the spin
polarization. It can be written as
\begin{equation}
\frac{1}{2}h_{s}^{z} = - \frac{i}{2}D \sum_{{\bf q} \eta} \sqrt{\frac{\hbar}{
  2 \rho \omega_{{\bf q}, \eta} {V} }} (\xi_{y \eta} q_{y} - \xi_{z \eta}
  q_{z}) q_{z} e^{i {\bf q}\cdot {\bf r}} \ .
\end{equation}
Comparing this to the electron-BP interaction Eq.\ (\ref{ep}),
the ratio is $\approx \hbar D q/\Xi$, which is about
$10^{-3}$. Therefore, the second-order term of the direct spin-phonon coupling due to
the phonon-induced strain together with the hyperfine interaction
 is very small and can be neglected. Also
the coaction of the $g$-factor fluctuation and the hyperfine interaction is
very small compared to that of the electron-BP interaction jointly with the
hyperfine interaction as $\mu_BB/\Xi$ is around $10^{-5}$ when $B=1$\ T. Therefore
it can also be neglected. In the following, we only
retain the first and the third order terms
$V_{eI-ph}^{(1)}$ and $V_{eI-ph}^{(3)}$ in calculating the spin 
relaxation time.

The spin dephasing time induced by the hyperfine interaction can be calculated
from the non-Markovian kinetic Eq.\
(\ref{hyp_eom2}), for unpolarized initial nuclear spin state $|n_0\rangle$,
resulting in
\begin{equation}
\langle S_{+}^{(n_0)} \rangle_{t} \propto \sum_{k}  f_{k}  A^{2} v_{0}^{2} \int d
  {\mathbf r} |\psi_k ( {\mathbf r} )|^{4}
\cos ( \frac{A v_{0}}{2} | \psi_k ({\mathbf r}) |^{2} t )\ ,
\label{dynamics}
\end{equation}
where $f_{k}$ is the thermo-equilibrium distribution of the orbital
degree of freedom. When only the lowest two Zeeman sublevels are
considered, assuming a simple form of the wavefunction,
$|\Psi(\mathbf{r})|^2=\frac{1}{a_{z} d_{\|}^2\pi} \exp(-r_{\|}^2/d_0^2)$ 
with $d_{\|}$/$a_{z}$ representing the
QD diameter/quantum well width, and $r_{\|}=x^2+y^2$,
the integration can be carried out:
\begin{equation}
\langle S_{+}^{(n_0)} \rangle_{t} \propto
  \frac{\cos(t/t_0)-1}{(t/t_0)^2} + \frac{\sin(t/t_0)}{t/t_0}\ .
\label{t0d}
\end{equation}
Here, $t_0=(2\pi a_{z} d_{\|}^2)/(A v_{0})$ determines the spin dephasing
time. Note that $t_0$ is proportional to the factor $a_{z} d_{\|}^2$
where $a_{z}$/$d_{\|}^2$ is the characteristic length/area of the
QD along $z$ direction / in the quantum well plane.  By solving Eq.\
(\ref{hyp_eom2}) for various $n$, and summing over $n$, we obtain 
$||\langle
  S_{+}\rangle_t||=\sum_{n} |\langle S_{+}^{(n)}\rangle_t|$. We then
define the time when the envelop of 
$||\langle  S_{+}\rangle_t||$ decays to
$1/e$ of its initial value as the spin dephasing time $T_{2}$. 
As mentioned above
the hyperfine interaction can not transfer an energy of the order of
the Zeeman splitting, thus the hyperfine interaction alone can not
lead to any spin relaxation.\cite{Witzel2}

In the above discussion, the nuclear spin dipole-dipole
 interaction is neglected. Recently, more careful examinations based
  on quantum cluster expansion method or pair correlation
  method have been performed.\cite{Witzel3,Yao,Witzel2,Witzel} 
  In these works, the nuclear spin dipole-dipole interaction is also
  included. This interaction together with the hyperfine mediated
  nuclear spin-spin interaction is the origin of the fluctuation of
  the nuclear spin bath. To the lowest order, the fluctuation is
  dominated by nuclear spin pair
  flips.\cite{Witzel3,Yao,Witzel2,Witzel} This fluctuation provides
  the source of the electron spin dephasing, as the electron spin is
  coupled to the nuclear spin system via hyperfine interaction. 
Our method used here includes only the hyperfine interaction to 
the second order in scattering. However, it is found that the
  dipole-dipole-interaction--induced spin dephasing is much weaker
  than the hyperfine interaction for a QD with $a=2.8$\ nm and $d_0$ =
  27\ nm until the parallel magnetic field is larger than $\sim 20$\
T.\cite{Yao}  Therefore, for the situation in this
paper, the nuclear dipole-dipole-interaction--induced spin
dephasing can be ignored.\cite{comment2}

\section{Spin decoherence due to various mechanisms}

Following the equation-of-motion approach developed in Sec.\ II, we perform
a numerical calculation of the spin  relaxation and
dephasing times in GaAs QDs. Two magnetic field configurations are
  considered: i.e.,  the magnetic fields perpendicular and
parallel to the well
  plane (along $x$-axis). The temperature is taken to be $T=4$\ K unless
otherwise specified. For all the cases we considered in this
manuscript, the orbital
 level splitting is larger than an energy corresponding to 40\  K. 
Therefore, the lowest Zeeman sublevels are mainly responsible for
 the spin decoherence. 
When calculating $T_1$, the initial
distribution is taken to be in the spin majority down state of the
eigen-state of the Hamiltonian $H_e$ with a
  Maxwell-Boltzmann distribution $f_{k}= C
  \exp[-\epsilon_{k}/(k_{B}T)]$ for different orbital levels 
($C$ is the normalization constant). For the calculation of
$T_2$, we assign the same distribution between different orbital
levels, but with a superposition of the two spin states within the
same orbital level. The parameters used in the 
calculation are listed in Table\ \ref{table1}.\cite{Paget, Madelung,Knap}

\begin{table}[htb]
\caption{Parameters used in the calculation}
\hspace{2cm}
\begin{tabular}{p{1cm} p{3.5cm} p{1cm} l }
\hline \hline
$\rho$ &  $5.3 \times 10^{3}$\ kg/m$^{3}$ & $\kappa$ & 12.9 \\
$v_{st}$  & $2.48 \times 10^{3}$\ m/s & $g$ & $-0.44$ \\
$v_{sl}$ & $5.29 \times 10^{3}$\ m/s & $\Xi$ & $7.0$\ eV \\
$e_{14}$ & $1.41 \times 10^{9}$\ V/m & $m^{\ast}$ & $0.067m_0$ \\
$A$    & $90$\ $\mu$eV & $A_{33}$ & $19.6$ \\
$\gamma_{0}$ & $27.5$\ \AA$^{3}\cdot$eV & $I$  & $\frac{3}{2}$  \\
$D$ & $1.59 \times 10^{4}$\ m/s &  $a_{0}$ &  $5.6534$\ \AA  \\
\hline \hline
\end{tabular}
\label{table1}
\end{table}

\subsection{Spin Relaxation Time $T_{1}$}

We now study the spin relaxation time and show how it changes
with the well width $a$, the magnetic field $B$ 
and the effective diameter $d_{0}=\sqrt{\hbar \pi/m^{\ast} \omega_{0}}$.
We also compare the relative contributions
from each relaxation mechanism.

\begin{figure}[bth]
  \begin{center}
\includegraphics[height=5.5cm]{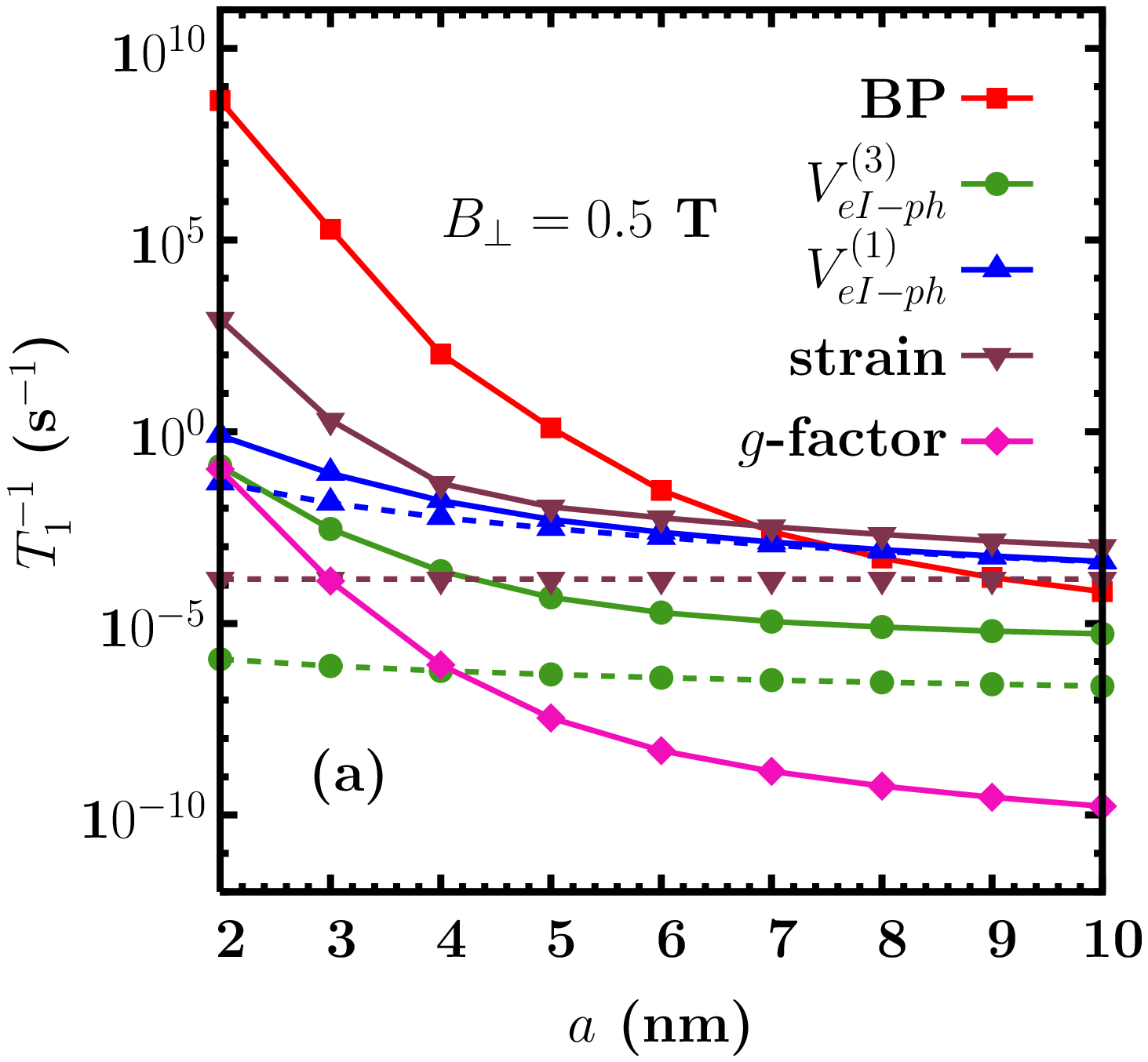}
\includegraphics[height=5.5cm]{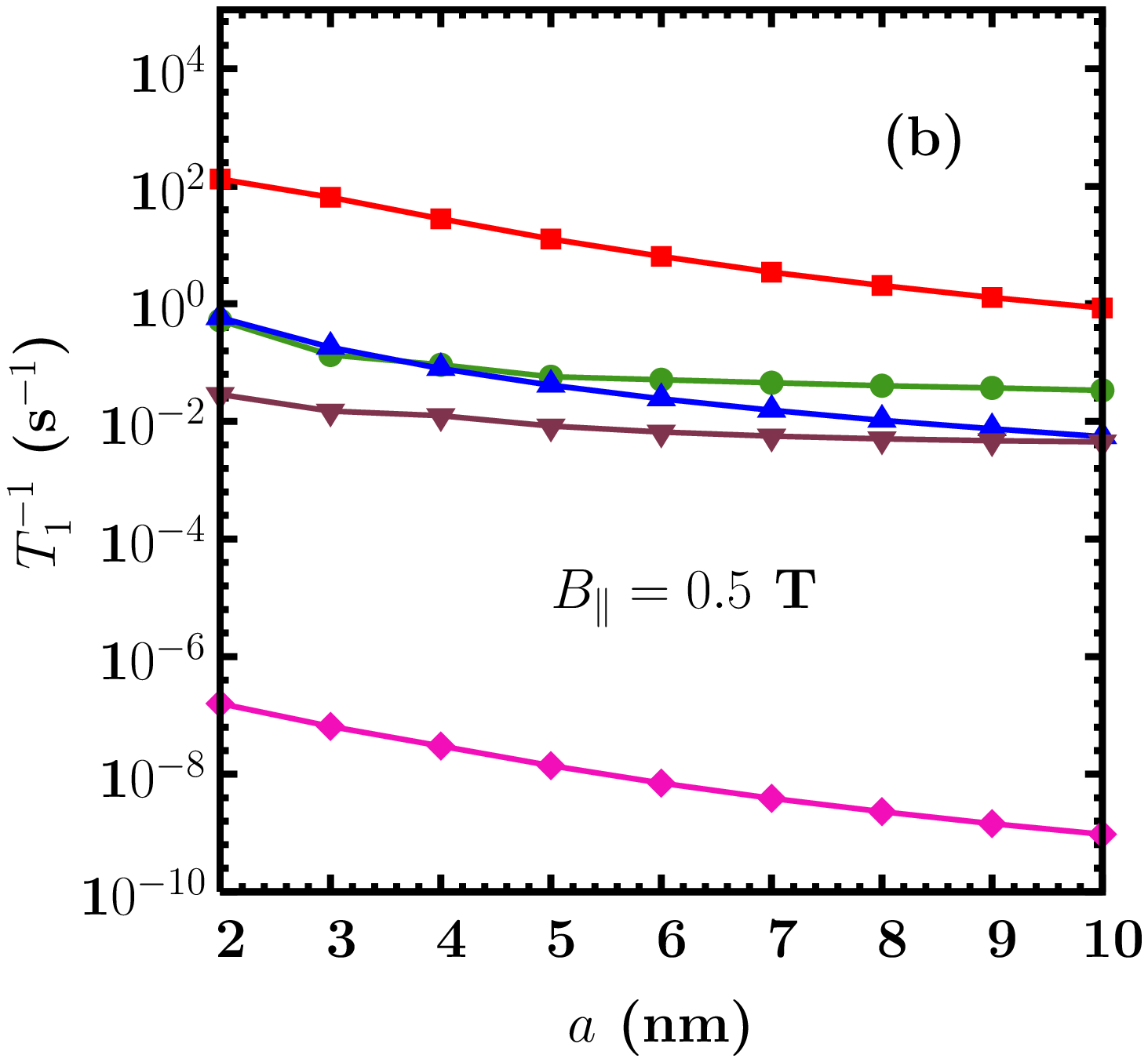}
\end{center}
\caption{(Color online) $T_{1}^{-1}$ induced by different mechanisms
  {\em vs.} the well width for (a): perpendicular magnetic 
  field $B_{\perp}=0.5$\ T with (solid curves) and without
  (dashed curves) the SOC; (b) parallel magnetic field $B_{\|}=0.5$\
  T with the SOC. The effective diameter $d_{0}=20$\ nm, and temperature $T=4$\ K.
Curves with $\blacksquare$ --- $T_{1}^{-1}$ induced by the electron-BP
scattering together with the SOC;
Curves with $\bullet$ --- $T_{1}^{-1}$ induced by the
  second-order process of the
  hyperfine interaction together with the BP ($V_{eI-ph}^{(3)}$);
Curves with
  $\blacktriangle$ --- $T_{1}^{-1}$
 induced by the first-order process of the hyperfine interaction
together with the BP ($V_{eI-ph}^{(1)}$);
Curves with $\blacktriangledown$ --- $T_{1}^{-1}$ induced by the
 direct spin-phonon coupling due to phonon-induced strain; Curves
with $\blacklozenge$ --- $T_{1}^{-1}$ induced by the $g$-factor fluctuation.}
\label{fig1}
\end{figure}

\subsubsection{Well width dependence}

In Fig.\ 1(a) and (b), the spin relaxation times induced by
different mechanisms are plotted as function of the width of the
quantum well in which the QD is confined for perpendicular magnetic
field $B_{\perp}=0.5$\ T and parallel magnetic field $B_{\|}=0.5$\ T
respectively. We first concentrate on the perpendicular
 magnetic field case. In Fig.\ 1(a), the 
 calculation indicates that the spin relaxation 
  due to each mechanism decreases with the increase of well width. 
Particularly the electron-BP scattering mechanism decreases
 much faster than the other mechanisms. 
It is indicated in the figures that when the well width is small (smaller
than 7\ nm in the present case), the spin relaxation time is determined
by the electron-BP scattering together with the SOC. However, for  wider
well widths, the direct spin-phonon coupling due to phonon-induced strain and the
first-order process of hyperfine interaction combined with the
electron-BP scattering becomes more important. 
The decrease of spin relaxation due to each mechanism is mainly
 caused by the decrease of the SOC which is proportional to $a^{-2}$. The SOC
 has two effects which are crucial. First, in the second order
 perturbation the SOC contributes a finite correction to the Zeeman
 splitting which determines the absorbed/emitted phonon frequency
 and wave-vector.\cite{Cheng} Second, it leads to spin mixing. 
 The decrease of the SOC thus leads to the decrease of
  Zeeman splitting and  spin mixing. The former leads to small
  phonon wave-vector and small phonon absorption/emission
 efficiency.\cite{Cheng} Therefore the electron-BP mechanism decreases
rapidly with increasing $a$. On the other hand, 
the other two largest mechanisms can flip spin without the help of
the SOC. The spin relaxations due to these two mechanisms decrease in a
relatively mild way. It is further confirmed that without
 SOC they decreases in
  a much milder way with increasing $a$ (dashed curves in Fig.\ 1).
It is also  noted that the spin relaxation rate due to
the $g$-factor fluctuation is at least six orders of magnitude smaller than
that due to the leading spin decoherence mechanisms and can therefore
be neglected.

It is noted that in the calculation, the SOC is
always included as it has large effect on the eigen-energy and
eigen-wavefunction of the electron.\cite{Cheng}
We also show the spin relaxation times induced by the
hyperfine interactions ($V^{(1)}_{eI-ph}$ and $V^{(3)}_{eI-ph}$)
and the direct spin-phonon coupling due to the phonon-induced
strain but without the SOC as in the
literature.\cite{Erlingsson,Abalmassov,Kim}
It can be seen clearly that the spin relaxation that includes the SOC
is {\em much} larger than that without the SOC. For example,
 the spin relaxation induced by the second-order process of
the hyperfine interaction together
 with the BP ($V_{eI-ph}^{(3)}$) is at least one order of magnitude
larger when the SOC is included than that when the SOC is
neglected.
This is because when the SOC is neglected,
 $\langle m|H_{eI} ({\mathbf r})
 |\ell_{1}\rangle$ and $\langle \ell_{2}| H_{eI} ({\mathbf r}) |m\rangle$ in
Eq.\ (\ref{two-order}) are small as
the matrix elements of $H_{eI}({\mathbf r})$
between different orbital energy levels are very small. However, when
the SOC is taken into account, the spin-up and -down
levels with different orbital quantum numbers are mixed
and therefore $|\ell\rangle$ and $|m\rangle$ include the components
with the same orbital quantum number.
Consequently the  matrix elements of $\langle m|H_{eI} ({\mathbf r})
|\ell_{1}\rangle$ and $\langle \ell_{2}| H_{eI} ({\mathbf r})
|m\rangle$ become much larger. Therefore, spin relaxation induced by
this mechanism depends crucially on the SOC.

It is emphasized from the above discussion that
the SOC  should be included
in each spin relaxation mechanism. In the following calculations it is always
included unless otherwise specified. 
In particular in reference to the mechanism of
electron-BP interaction, we always consider it 
together with the SOC.

We further discuss the parallel magnetic field case. In Fig.\ 1(b) the
spin relaxation times due to different mechanisms are plotted as
function of the quantum well width for same parameters as Fig.\ 1(a), but
with a parallel magnetic field $B_{\|}=0.5$\ T. It is noted that
the spin relaxation rate due to each mechanism becomes much smaller
for small $a$ compared with the perpendicular case. Another feature is
that the spin relaxation due to each mechanism decrease in a much slower
rate with increasing $a$. The electron-BP mechanism is dominant
even at $a=10$\ nm but decrease faster than other mechanisms with
$a$. It is expected to be less effective than the 
$V_{eI-ph}^{(3)}$ mechanism or $V_{eI-ph}^{(1)}$ mechanism or the
direct spin-phonon coupling due to phonon-induced strain mechanism for
large enough $a$. The $g$-factor 
fluctuation mechanism is negligible again. These features can be
explained as follows. For parallel magnetic field the contribution of
the SOC to Zeeman splitting is much less than in the perpendicular
magnetic field geometry.\cite{Destefani} Moreover, this contribution is
{\em negative} which makes Zeeman splitting smaller.\cite{Destefani}
Therefore, the phonon absorption/emission efficiency becomes much
smaller for small $a$, {\em i.e.}, large SOC. When $a$ increases, the
Zeeman splitting increases. However, the spin mixing decreases. The
former effect is weak, and only cancels part of the latter, thus the
spin relaxation due to each mechanism decrease slowly with $a$.

\begin{figure}[bth]
\begin{center}
\includegraphics[height=5.5cm]{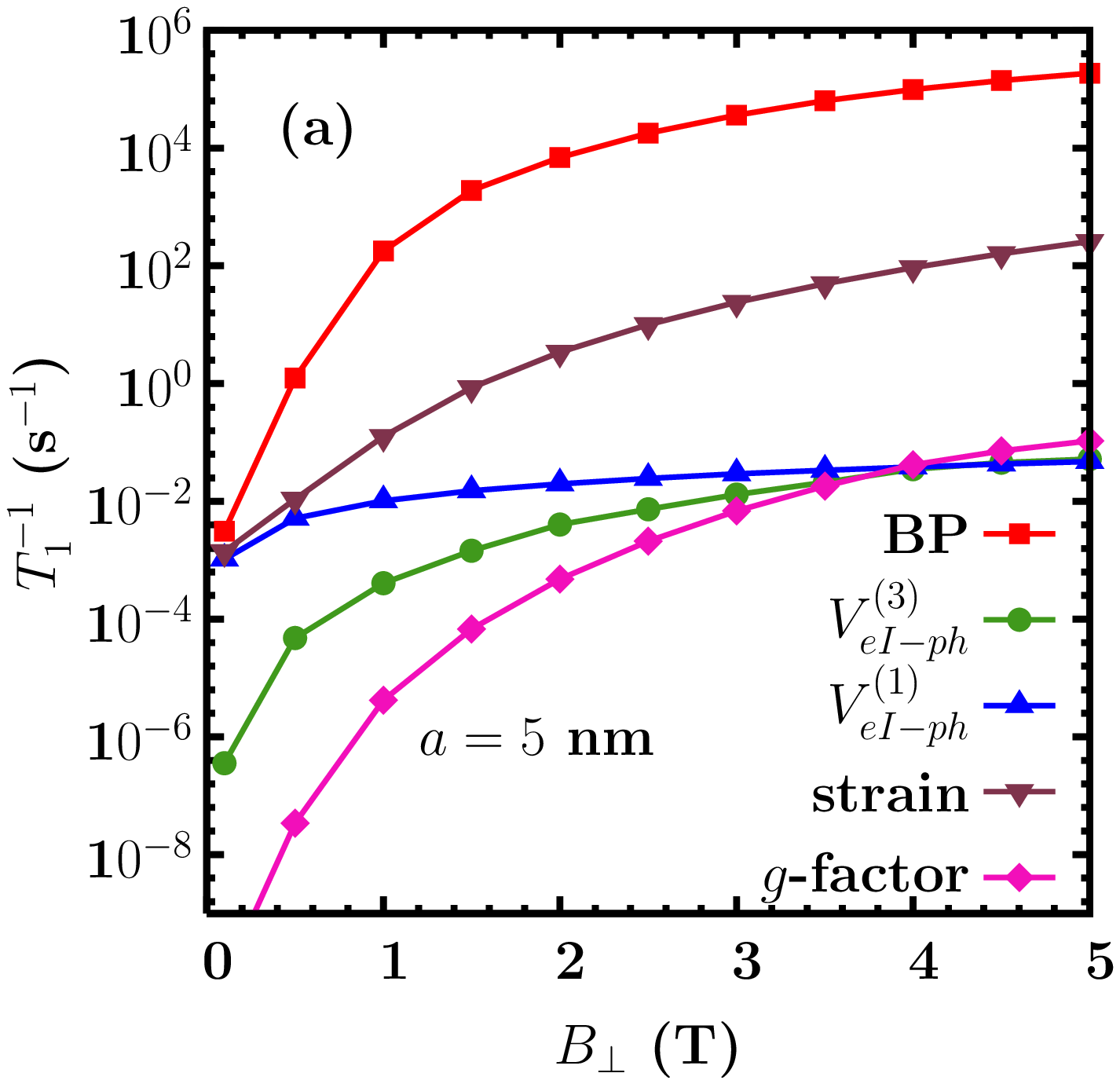}
\includegraphics[height=5.5cm]{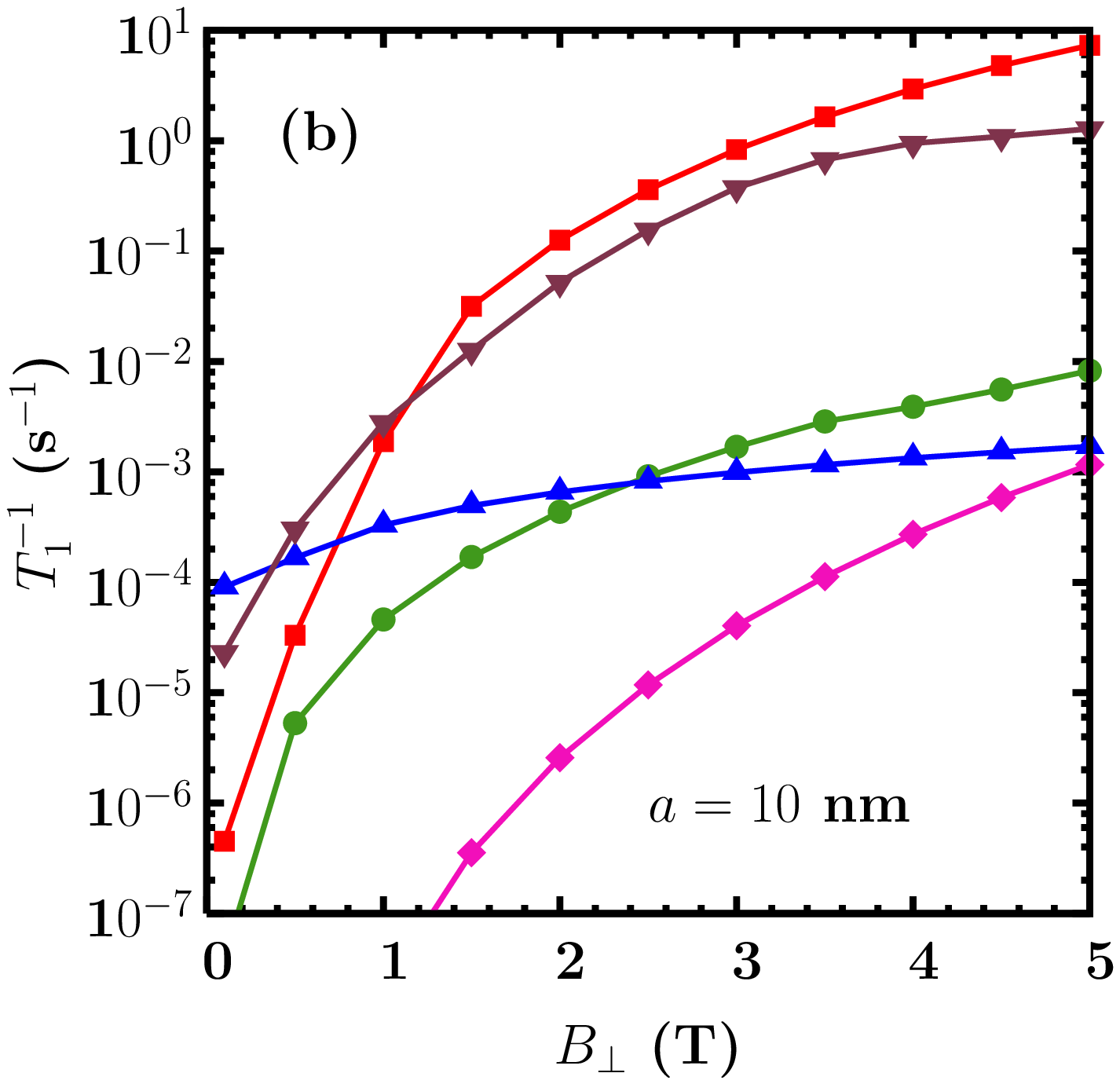}
\end{center}
\caption{(Color online) $T_{1}^{-1}$ induced by different mechanisms
{\em vs}. the perpendicular magnetic field $B_{\perp}$ for $d_{0}=20$\ nm and
(a) $a=5$\ nm  and  (b) $10$\ nm. $T=4$\ K.
Curves with $\blacksquare$ --- $T_{1}^{-1}$ induced by the electron-BP
scattering;
Curves with $\bullet$ --- $T_{1}^{-1}$ induced by the
  second-order process of the
  hyperfine interaction together with the BP ($V_{eI-ph}^{(3)}$);
Curves with  $\blacktriangle$ --- $T_{1}^{-1}$
 induced by the first-order process of the hyperfine interaction
together with the BP ($V_{eI-ph}^{(1)}$);
Curves with $\blacktriangledown$ --- $T_{1}^{-1}$ induced by the
 direct spin-phonon coupling due to phonon-induced strain; Curves
with $\blacklozenge$ --- $T_{1}^{-1}$ induced by the $g$-factor fluctuation.}
\label{fig2}
\end{figure}

\begin{figure}[thb]
\begin{center}
\includegraphics[height=5.5cm]{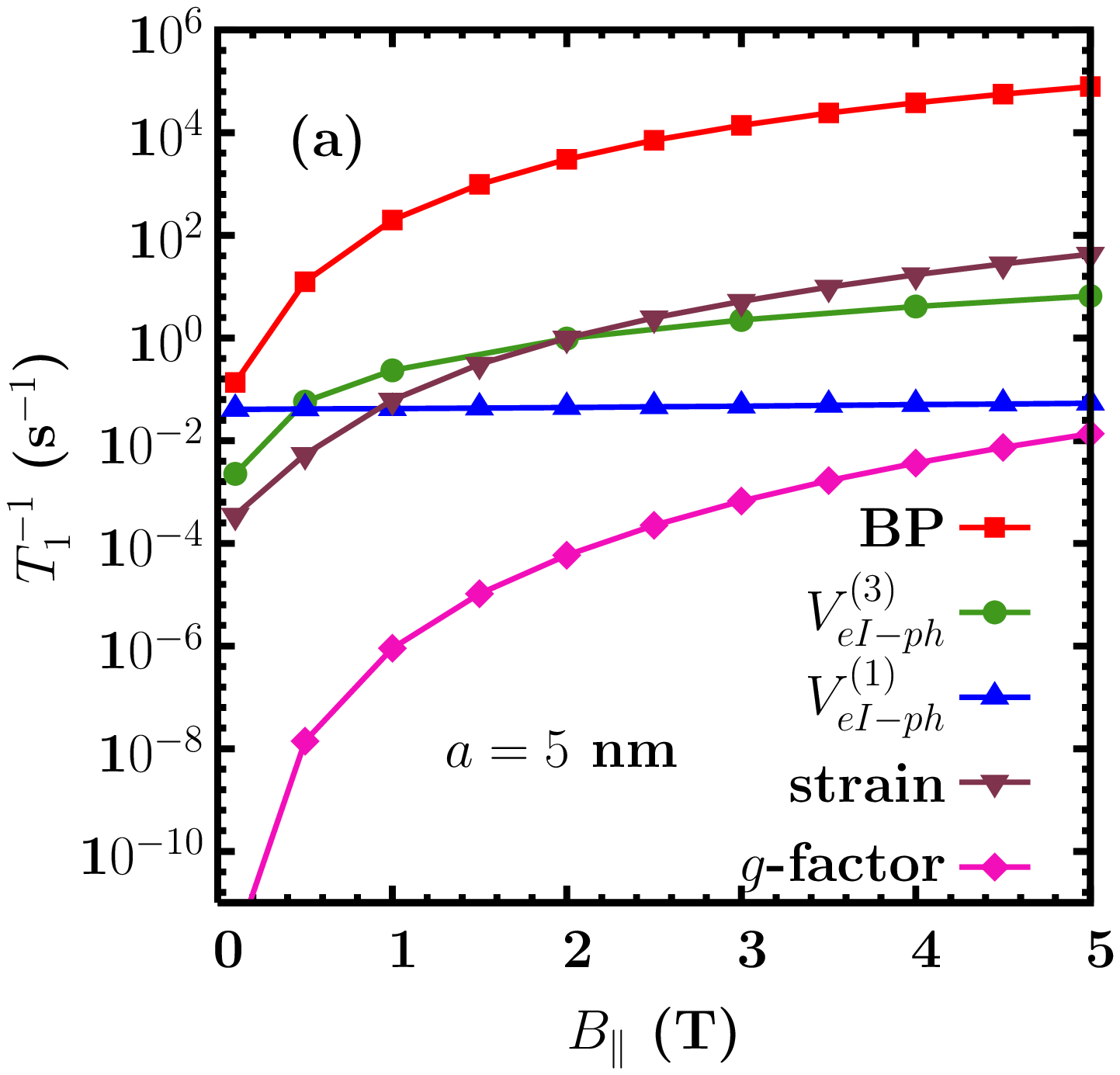}
\includegraphics[height=5.5cm]{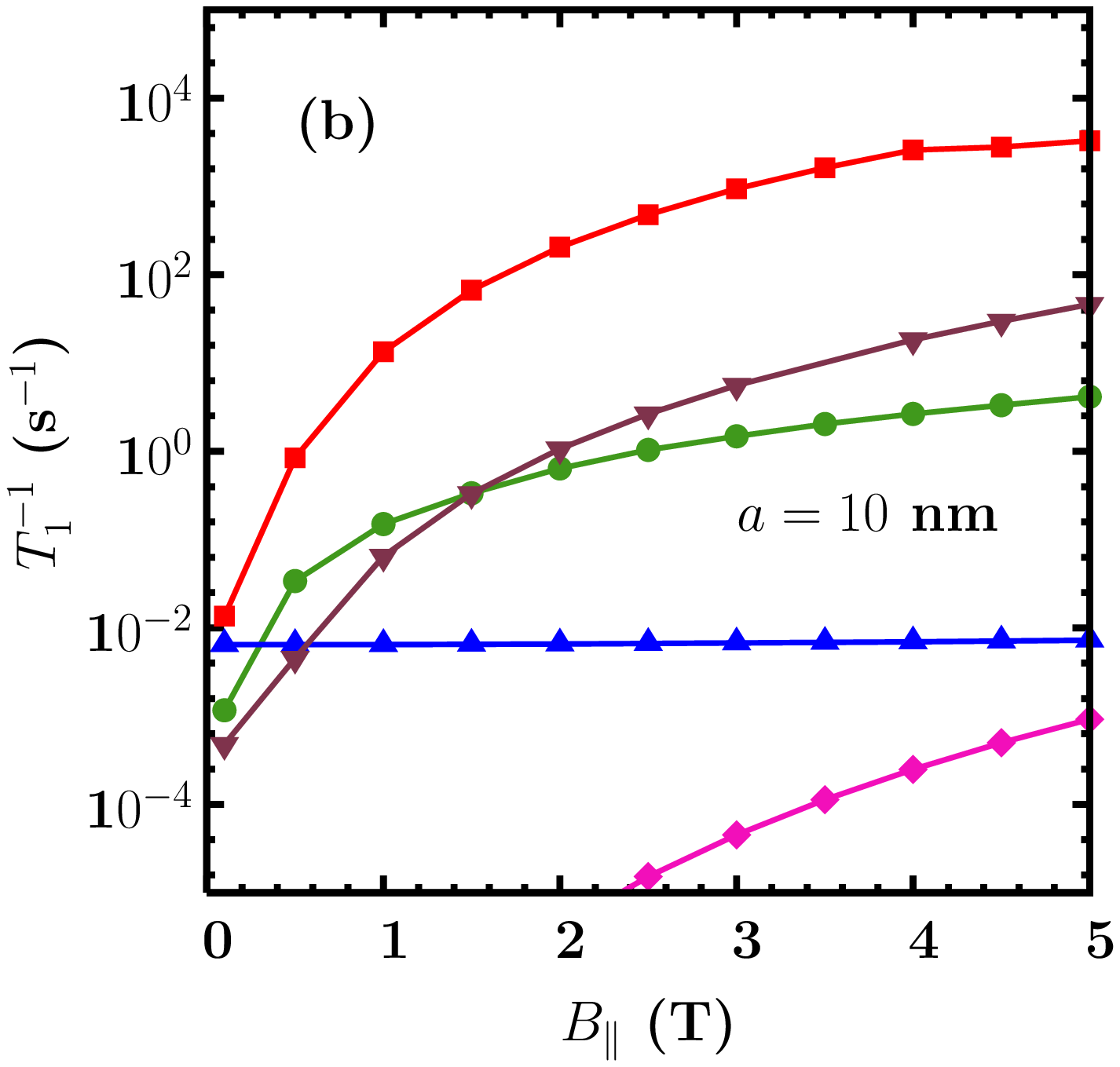}
\end{center}
\caption{(Color online) $T_{1}^{-1}$ induced by different mechanisms
{\em vs}. the parallel magnetic field $B_{\|}$ for $d_{0}=20$\ nm and
(a) $a=5$\ nm  and  (b) $10$\ nm. $T=4$\ K.
Curves with $\blacksquare$ --- $T_{1}^{-1}$ induced by the electron-BP
scattering; 
Curves with $\bullet$ --- $T_{1}^{-1}$ induced by the
  second-order process of the
  hyperfine interaction together with the BP ($V_{eI-ph}^{(3)}$);
Curves with  $\blacktriangle$ --- $T_{1}^{-1}$
 induced by the first-order process of the hyperfine interaction
together with the BP ($V_{eI-ph}^{(1)}$);
Curves with $\blacktriangledown$ --- $T_{1}^{-1}$ induced by the
 direct spin-phonon coupling due to phonon-induced strain; Curves
with $\blacklozenge$ --- $T_{1}^{-1}$ induced by the $g$-factor fluctuation.}
\label{fig3}
\end{figure}

\subsubsection{Magnetic Field Dependence}

We first study the perpendicular-magnetic-field
case. The magnetic field dependence of $T_{1}$ for two
different well widths are shown in Fig.\ \ref{fig2}(a) and Fig.\
\ref{fig2}(b). In the calculation, $d_{0}=20$\ nm. 
It can be seen that the effect of each mechanism increases
 with the magnetic field. Particularly the
 electron-BP mechanism increases much faster than other ones and becomes 
 dominant at high magnetic fields. 
For small well width (5\ nm in Fig.\ \ref{fig2}a),
the spin relaxation induced by the
electron-BP scattering is dominant
except at  very low magnetic fields (0.1\ T in the figure) where
contributions from the first-order process of hyperfine
  interaction together with the electron-BP scattering and 
the direct spin-phonon coupling due to
phonon-induced strain also contribute.
It is interesting to see that when $a$ is increased to $10$\ nm,
the electron-BP scattering  is the largest
spin relaxation mechanism only at high magnetic
fields ($>$1.1\ T). For 0.4\ T
$<B_{\perp}<1.1$\ T ($B_{\perp}<0.4$\ T), the direct spin-phonon coupling due to 
the phonon-induced
strain (the first order hyperfine interaction together with the
BP ) becomes the largest relaxation mechanism.
It is also noted that
there is no single mechanism which dominates the whole spin relaxation.
Two or three mechanisms are jointly responsible for the spin relaxation.
It is indicated that the spin relaxations induced by different
 mechanisms all increase with $B_{\perp}$. This can be understood  from
a perturbation theory: when the magnetic field is small the spin
relaxation between two Zeeman split states for each mechanism is proportional to
$\bar{n}(\Delta E)(\Delta E)^m$ ($\Delta E$ is the Zeeman
splitting) with $m=7$ for electron-BP scattering
due to the deformation potential\cite{Cheng,Stano2}
 and for the second-order process of the
 hyperfine interaction together with the electron-BP scattering
due to the deformation potential $V_{eI-ph}^{(3)}$;\cite{Erlingsson} $m=5$
 for electron-BP scattering due to the piezoelectric coupling\cite{Nazarov,Cheng,Stano2}
 and for the second-order process of the
 hyperfine interaction together with the electron-BP scattering
 due to the piezoelectric coupling  $V_{eI-ph}^{(3)}$;\cite{Erlingsson} and
 $m=5$ for the direct
 spin-phonon coupling due to phonon-induced strain;\cite{Nazarov} $m=1$ for the first-order
 process of the hyperfine interaction together with the BP
 $V_{eI-ph}^{(1)}$. The spin relaxation induced by the $g$-factor
 fluctuation is proportional to $\bar{n}(\Delta E)(\Delta E)^5
 B_{\perp}^{2}$. For most of the cases
studied, $\Delta E$ is smaller than $k_{B}T$, hence $\bar{n}(\Delta E)
\sim k_BT/\Delta E$, and $\bar{n}(\Delta E)(\Delta E)^m\sim (\Delta E)^{m-1}$. $m>1$ hold for
all mechanism except the $V_{eI-ph}^{(1)}$ mechanism, therefore
the spin relaxation due to these mechanisms increases with
increasing $B_{\perp}$. However, from Eq.\ (\ref{1st-order}) one can see that
 it has a term with ${\mathbf \nabla}_{{\mathbf r}}$, which indicates that
  the effect of this mechanism is proportional to $1/d_0$. As the
 vector potential of the magnetic field increases the confinement of
 the QD and gives rise to smaller effective diameter $d_0$, this mechanism
 also increases with the magnetic field in the perpendicular
  magnetic field geometry.

\begin{figure}[bth]
\begin{center}
\includegraphics[height=5.5cm]{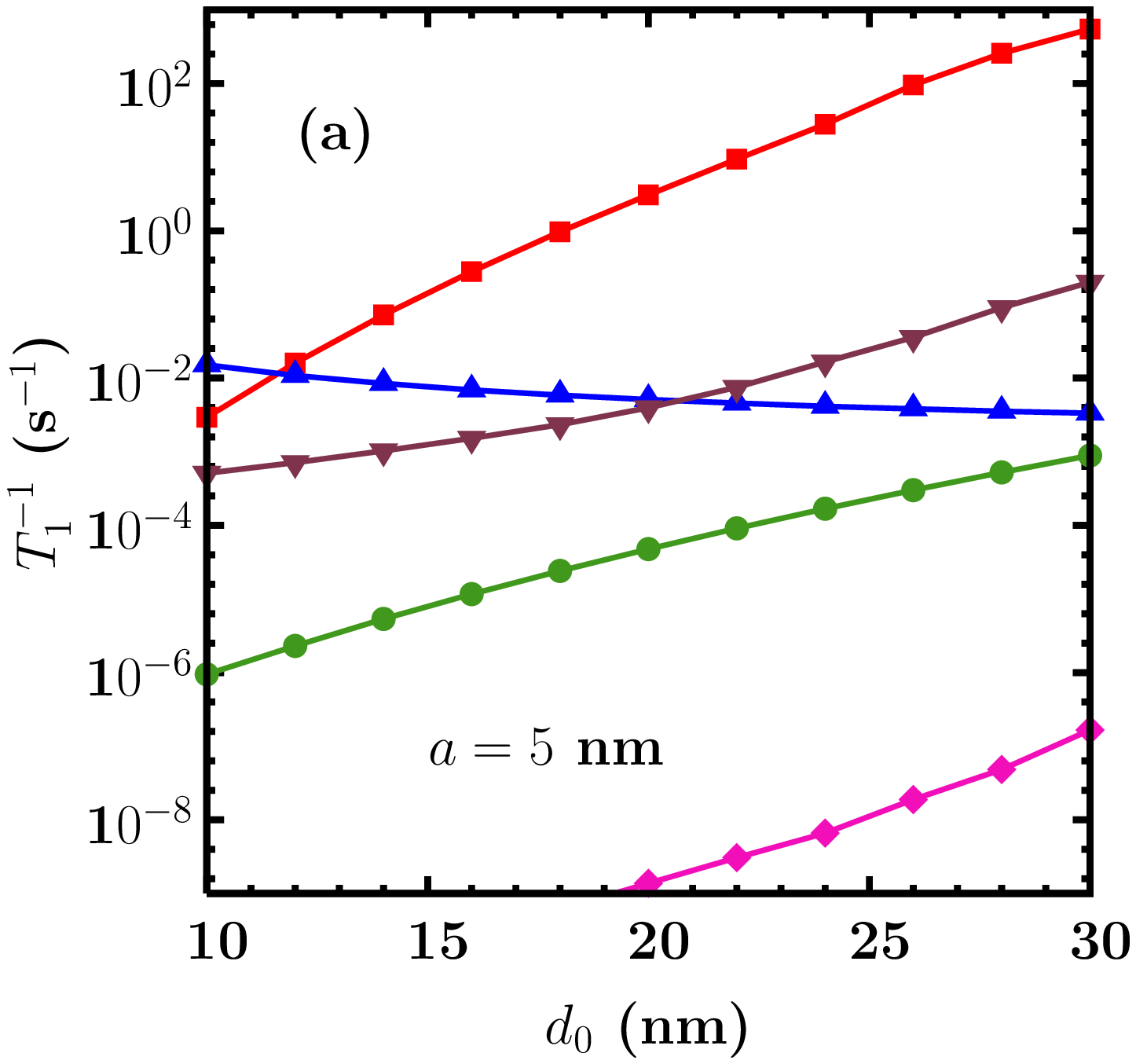}
\includegraphics[height=5.5cm]{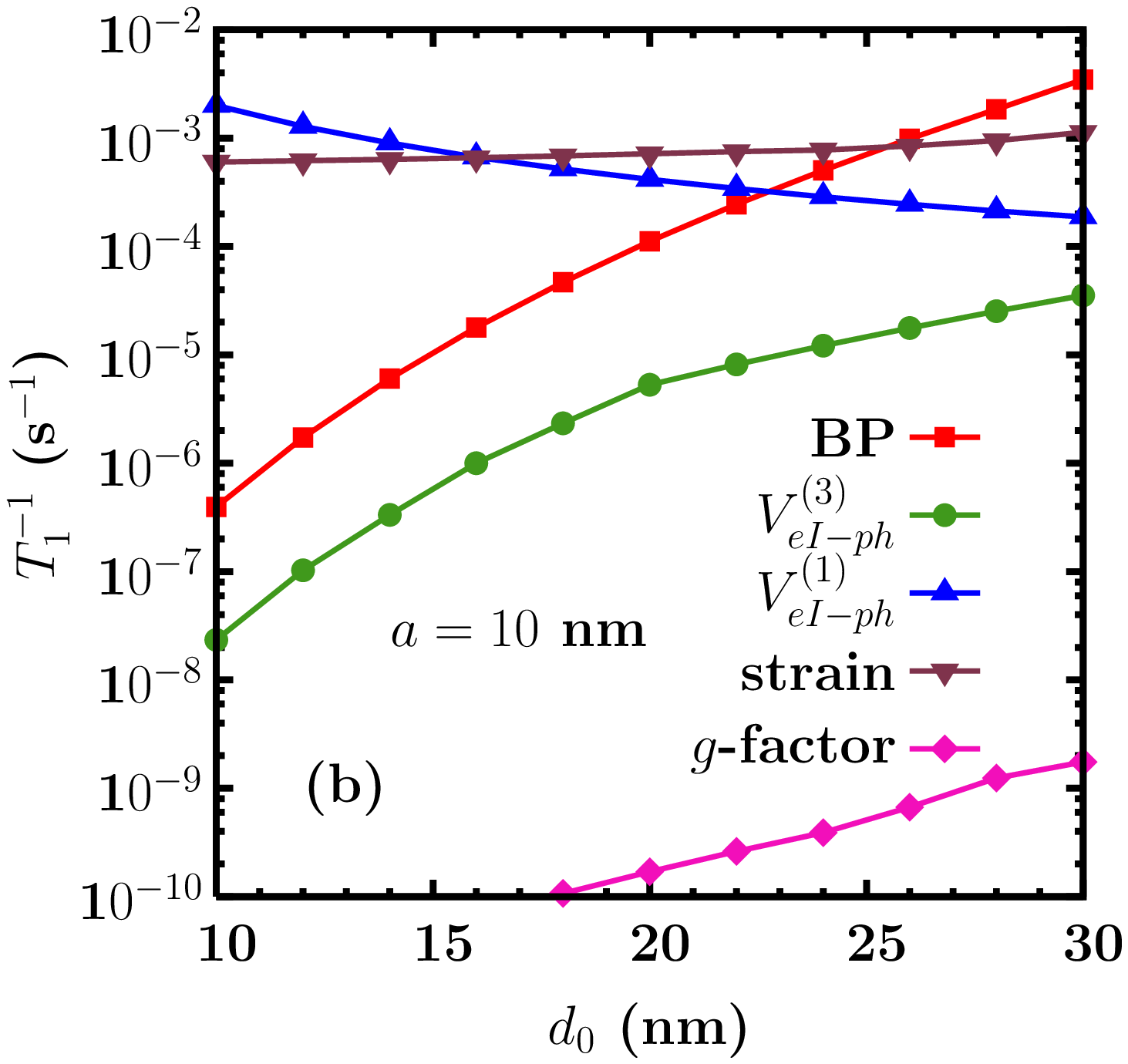}
\end{center}
\caption{(Color online) $T_{1}^{-1}$ induced by different mechanisms
{\em vs}. the effective diameter $d_{0}$ for $B_{\perp}=0.5$\ T and
(a) $a=5$\ nm  and (b) $10$\ nm. $T=4$\ K.
Curves with $\blacksquare$ --- $T_{1}^{-1}$ induced by the electron-BP
scattering; Curves with $\bullet$ -- $T_{1}^{-1}$ induced by the
 second-order process of the
hyperfine interaction together with the BP ($V_{eI-ph}^{(3)}$);
Curves with  $\blacktriangle$ --- $T_{1}^{-1}$
 induced by the first-order process of the hyperfine interaction
together with the BP ($V_{eI-ph}^{(1)}$);
Curves with $\blacktriangledown$ --- $T_{1}^{-1}$ induced by the
 direct spin-phonon coupling due to phonon-induced strain; Curves
with $\blacklozenge$ --- $T_{1}^{-1}$ induced by the $g$-factor fluctuation.}
\label{fig4}
\end{figure}

\begin{figure}[bth]
\begin{center}
\includegraphics[height=5.5cm]{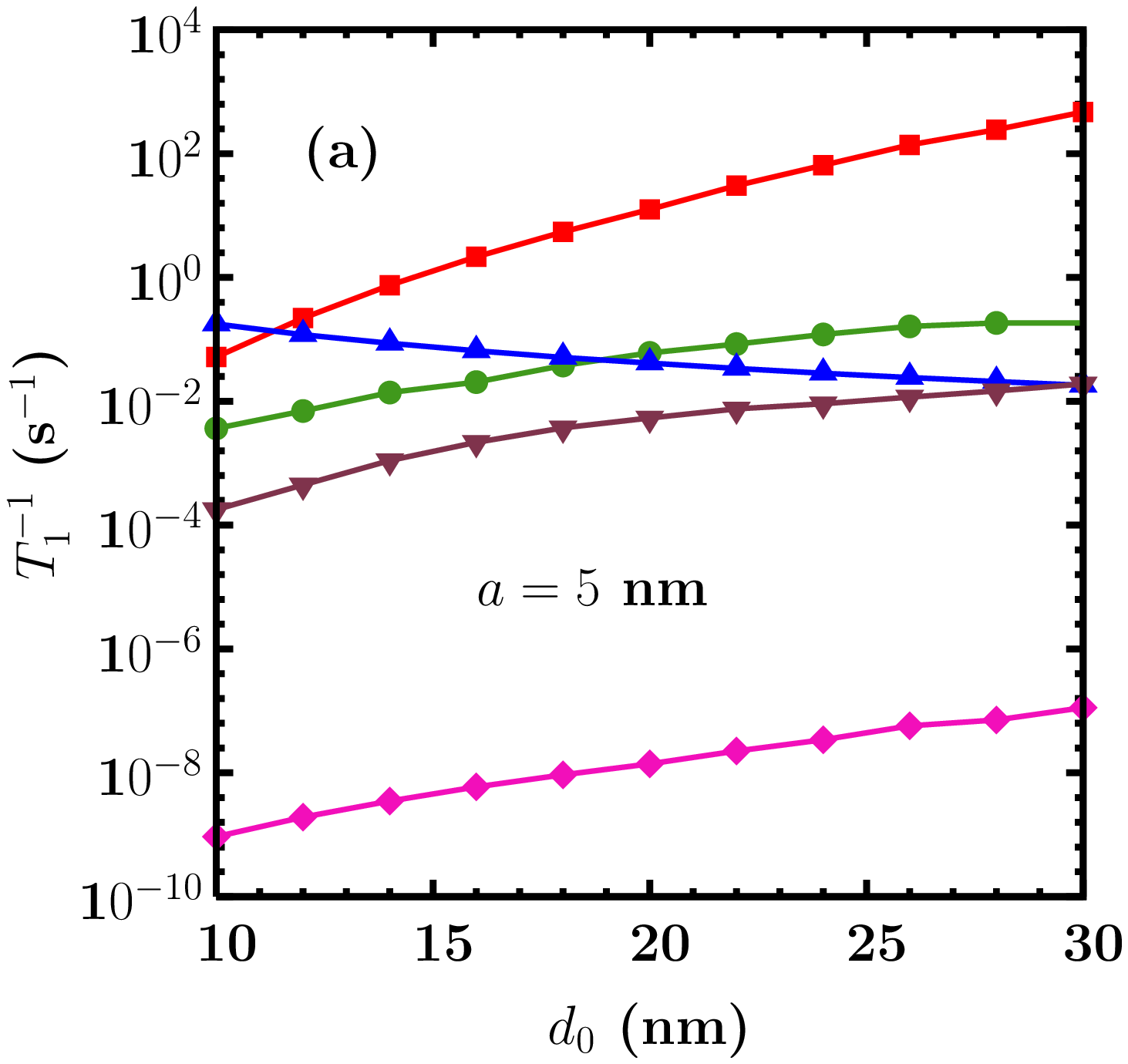}
\includegraphics[height=5.5cm]{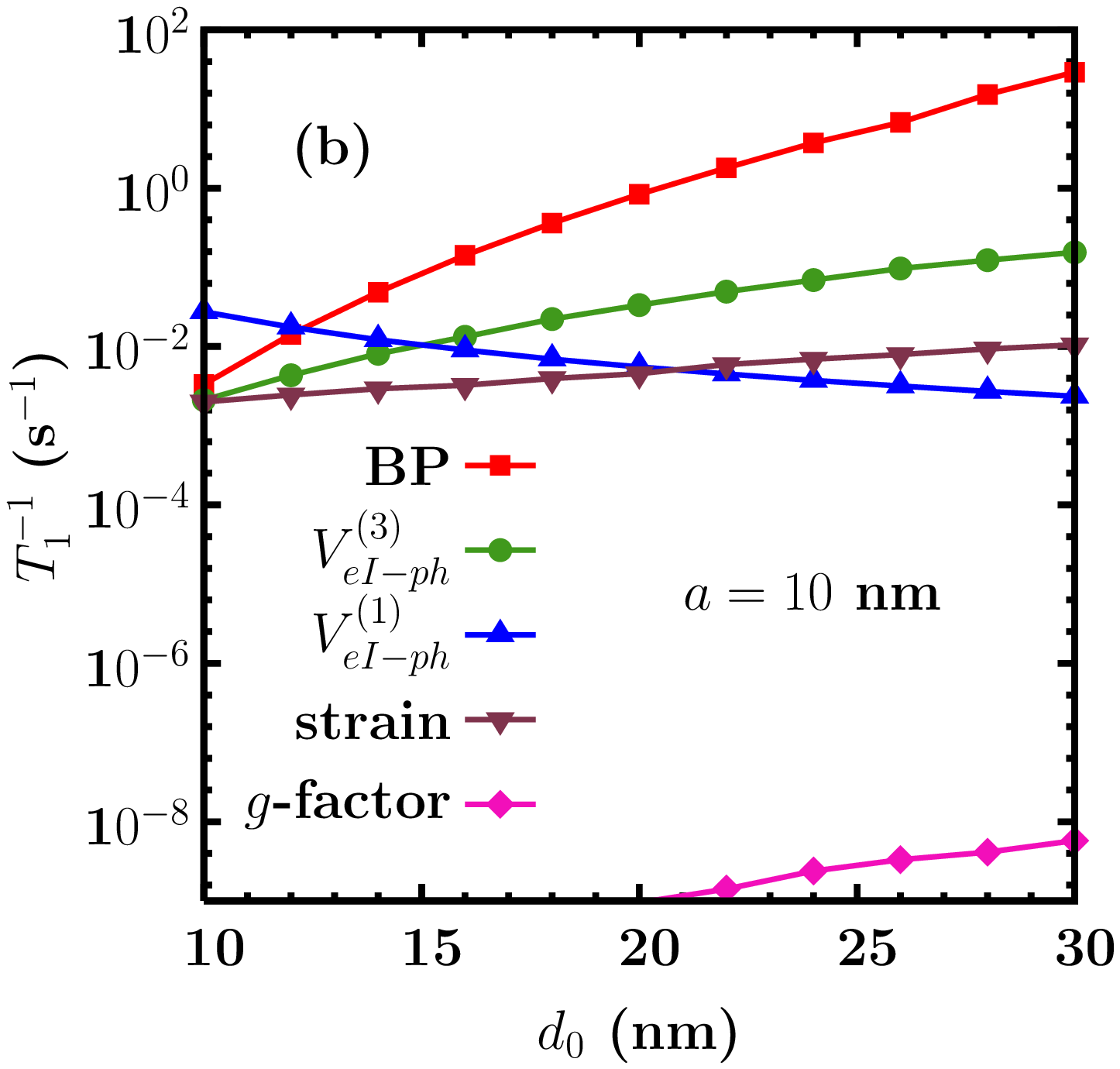}
\end{center}
\caption{(Color online) $T_{1}^{-1}$ induced by different mechanisms
{\em vs}. the effect diameter $d_{0}$ with $B_{\|}=0.5$\ T and
(a) $a=5$\ nm  and  (b) $10$\ nm. $T=4$\ K.
Curves with $\blacksquare$ --- $T_{1}^{-1}$ induced by the electron-BP
scattering;
Curves with $\bullet$ --- $T_{1}^{-1}$ induced by the
  second-order process of the
  hyperfine interaction together with the BP ($V_{eI-ph}^{(3)}$);
Curves with  $\blacktriangle$ --- $T_{1}^{-1}$
 induced by the first-order process of the hyperfine interaction
together with the BP ($V_{eI-ph}^{(1)}$);
Curves with $\blacktriangledown$ --- $T_{1}^{-1}$ induced by the
 direct spin-phonon coupling due to phonon-induced strain; Curves
with $\blacklozenge$ --- $T_{1}^{-1}$ induced by the $g$-factor fluctuation.}
\label{fig5}
\end{figure}

We then study the case with the magnetic field parallel to the quantum
well plane. In Fig.\ \ref{fig3} the spin relaxation induced by different mechanisms
are plotted as function of the parallel magnetic field $B_{\|}$ for two
different well widths. In the calculation, $d_{0}=20$\ nm. It can be
seen that, similar to the case with perpendicular magnetic field, the
effects of most mechanisms increase with the magnetic field.
Also the electron-BP mechanism increases much 
faster than the other ones and becomes 
 dominant at high magnetic fields. However, without the orbital
 effect of the magnetic field in the present configuration, the
 effect of $V_{eI-ph}^{(1)}$ changes very little with the magnetic field.
 For both small ($5$\ nm in Fig.\ \ref{fig3}(a))  
and large ($10$\ nm in Fig.\ \ref{fig3}(b)) well widths, the
electron-BP scattering is dominant except at very low magnetic field
($0.1$\ T in the figure), where the first-order process of the hyperfine
interaction together with the electron-BP interaction $V_{eI-ph}^{(1)}$ also
contributes. 

\subsubsection{Diameter Dependence}

We now turn to the investigation of the diameter dependence of the
spin relaxation. We first concentrate on the perpendicular magnetic
field geometry. The spin relaxation rate due to each mechanism is
shown in Fig.\ \ref{fig4}a for a small ($a=5$\ nm) and Fig.\
\ref{fig4}b for a large ($a=10$\ nm) well widths respectively
with a fixed perpendicular magnetic field $B_{\perp}=0.5$\ T.
In the figure, the spin relaxation rate due each mechanism except 
$V_{eI-ph}^{(1)}$ increases with the effective diameter. Specifically,
the effect of the electron-BP mechanism increases very fast, while the
effect of the direct spin-phonon coupling due to phonon-induced strain
mechanism increases very mildly. The $V_{eI-ph}^{(1)}$ decreases with
$d_0$ slowly. Other mechanisms are unimportant. 
The electron-BP mechanism eventually dominates spin relaxation when the
diameter is large enough. The threshold increases from 12\ nm to 26\
nm when the well width increases from 5\ nm to 10\ nm. For small
diameter the $V_{eI-ph}^{(1)}$ and the direct spin-phonon coupling due
to phonon-induced strain mechanism dominate the spin relaxation.
The increase/decrease of the spin relaxation due to these mechanisms
can be understood from the following. The effect of the SOC on
the Zeeman splitting is proportional to $d_0^2$ for small magnetic
field.\cite{Cheng} The increase of $d_0$ thus leads to a
increase of Zeeman splitting, therefore the efficiency of the phonon
absorption/emission increases. Another effect is that the increase of
$d_0$ will increase the phonon absorption/emission efficiency due to
the increase of the form factor.\cite{Cheng} Thus the spin relaxation increases. 
Moreover, the spin mixing is also proportional to $d_0$ also.\cite{Cheng} This
leads to much faster increasing of the effect of the electron-BP
mechanism and the $V_{eI-ph}^{(3)}$ mechanism. 
However, the spin relaxation due to $V_{eI-ph}^{(1)}$ 
decreases with the diameter. This is because $V_{eI-ph}^{(1)}$
contains a term ${\mathbf \nabla}_{{\mathbf r}}$ [Eq.\
(\ref{1st-order})] which decreases with the increase of
$d_{0}$. Physically speaking, the decrease of the effect of
$V_{eI-ph}^{(1)}$ is due to the fact that the spin mixing due to the
hyperfine interaction decreases with the increase of the number of
nuclei within the dot $N$ as the random Overhauser field is proportional
to $1/\sqrt{N}$. The spin relaxation induced by the
$g$-factor is also negligible here for both small and large well width. 

We then turn to the parallel magnetic field case. In the calculation,
$B_{\|}=0.5$\ T. The results are shown for both small well width
($a=5$\ nm in Fig.\ \ref{fig5}(a)) and large well width ($a=10$\ nm in
Fig.\ \ref{fig5}(b)) respectively.
Similar to the perpendicular magnetic field case, 
the effect of every mechanism except the $V_{eI-ph}^{(1)}$ mechanism
increases with increasing diameter. The effect of the electron-BP
mechanism increases fastest and becomes dominant for $d_0>12$\ nm for
both small and large well width. For $d_0<12$\ nm for the two cases
the first-order process of the $V_{eI-ph}^{(1)}$ mechanism becomes
dominant. The effect of the $V_{eI-ph}^{(3)}$ mechanism become larger
than that of the direct spin-phonon coupling due to phonon-induced
strain mechanism. However, these two mechanism are still unimportant
and becomes more and more unimportant for larger $d_0$. Here, the spin
relaxation induced by the $g$-factor is negligible. 

\subsubsection{Comparison with Experiment}

In this subsection, we apply our analysis to experiment data  in
Ref. \onlinecite{Elzerman}. We first show that our calculation is in
good agreement with the experimental results. Then we compare
contributions from different mechanisms to spin relaxation as function
of the magnetic field. In the calculation we choose the quantum dot diameter
$d_0=56$\ nm ($\hbar\omega_0=1.1$\ meV as in experiment). The quantum
well is taken to be an infinite-depth well with $a=13$\ nm. The
Dresselhaus SOC parameter $\gamma_0\langle k_z^2\rangle$ is taken to be 4.5\
meV$\cdot$\AA and  the Rashba SOC parameter is 
3.3\ meV$\cdot$\AA. $T=0$\ K as $k_BT\ll g\mu_BB$ in the
experiment. The magnetic field is applied parallel to the 
well plane in [110]-direction. The Dresselhaus cubic term
is also taken into consideration. All these parameters are the same with
(or close to) those used in Ref.\ \onlinecite{Stano} in which a
calculation based on the electron-BP scattering mechanism agrees well
with the experimental results. For this mechanism, we reproduce their
results. The spin relaxation time measured by the experiments (black
dots with error bar in the figure) almost coincide with the calculated
spin relaxation time due to the electron-BP scattering mechanism
(curves with $\blacksquare$ in the figure).\cite{note2} 
It is noted from the figure that other mechanisms are unimportant
for small magnetic field. However, for large magnetic field the effect
of the direct-spin phonon coupling due to phonon-induced strain 
becomes comparable with that of the electron-BP mechanism. At
$B_{\|}=10$\ T, the two differs by a factor of $\sim 5$. 

\begin{figure}[thb]
\begin{center}
\includegraphics[height=5.5cm]{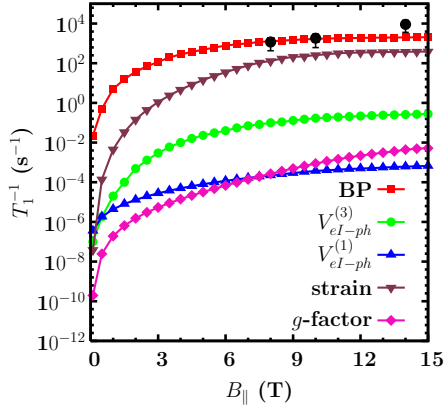}
\end{center}
\caption{(Color online) $T_{1}^{-1}$ induced by different mechanisms
{\em vs}. the parallel magnetic field $B_{\|}$ in the [110] direction
for $d_{0}=56$\ nm and $a=13$\ nm with both the Rashba and Dresselhaus
SOCs. $T=0$\ K. The black dots with 
error bar is the experimental results in Ref.\ \onlinecite{Elzerman}.
Curves with $\blacksquare$ --- $T_{1}^{-1}$ induced by the electron-BP
scattering; Curves with $\bullet$ --- $T_{1}^{-1}$ induced by the
second-order process of the hyperfine interaction together with the BP
($V_{eI-ph}^{(3)}$); Curves with  $\blacktriangle$ --- $T_{1}^{-1}$
induced by the first-order process of the hyperfine interaction
together with the BP ($V_{eI-ph}^{(1)}$); Curves with
$\blacktriangledown$ --- $T_{1}^{-1}$ induced by the direct
spin-phonon coupling due to phonon-induced strain; Curves 
with $\blacklozenge$ --- $T_{1}^{-1}$ induced by the $g$-factor fluctuation.}
\label{fig6}
\end{figure}

\subsection{Spin Dephasing Time $T_2$}

In this subsection, we investigate the spin dephasing time
for different well widths, magnetic fields and QD diameters. As in the
previous subsection, the contributions of the different mechanisms to
spin dephasing are compared.\cite{note1} To justify the first Born approximation in  
studying the hyperfine interaction induced spin dephasing, we focus mainly
on the high magnetic field regime of $B>3.5$\ T. A typical magnetic
field is $4$\ T. We also demonstrate via extrapolation that in the low
magnetic field regime spin dephasing is dominated by the hyperfine
interaction.

\subsubsection{Well Width Dependence}

\begin{figure}[bth]
 \begin{center}
\includegraphics[height=5.5cm]{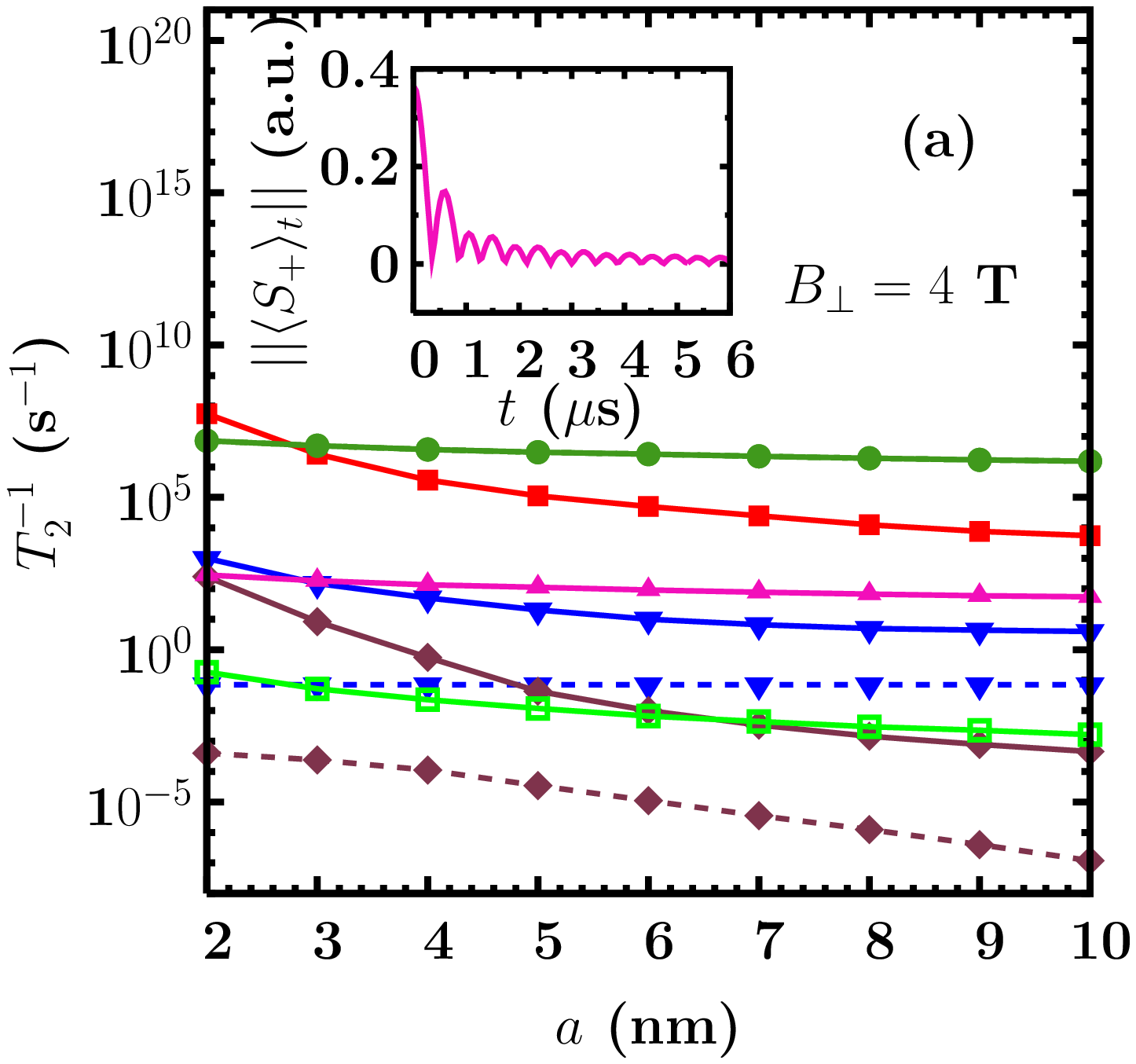}
\includegraphics[height=5.5cm]{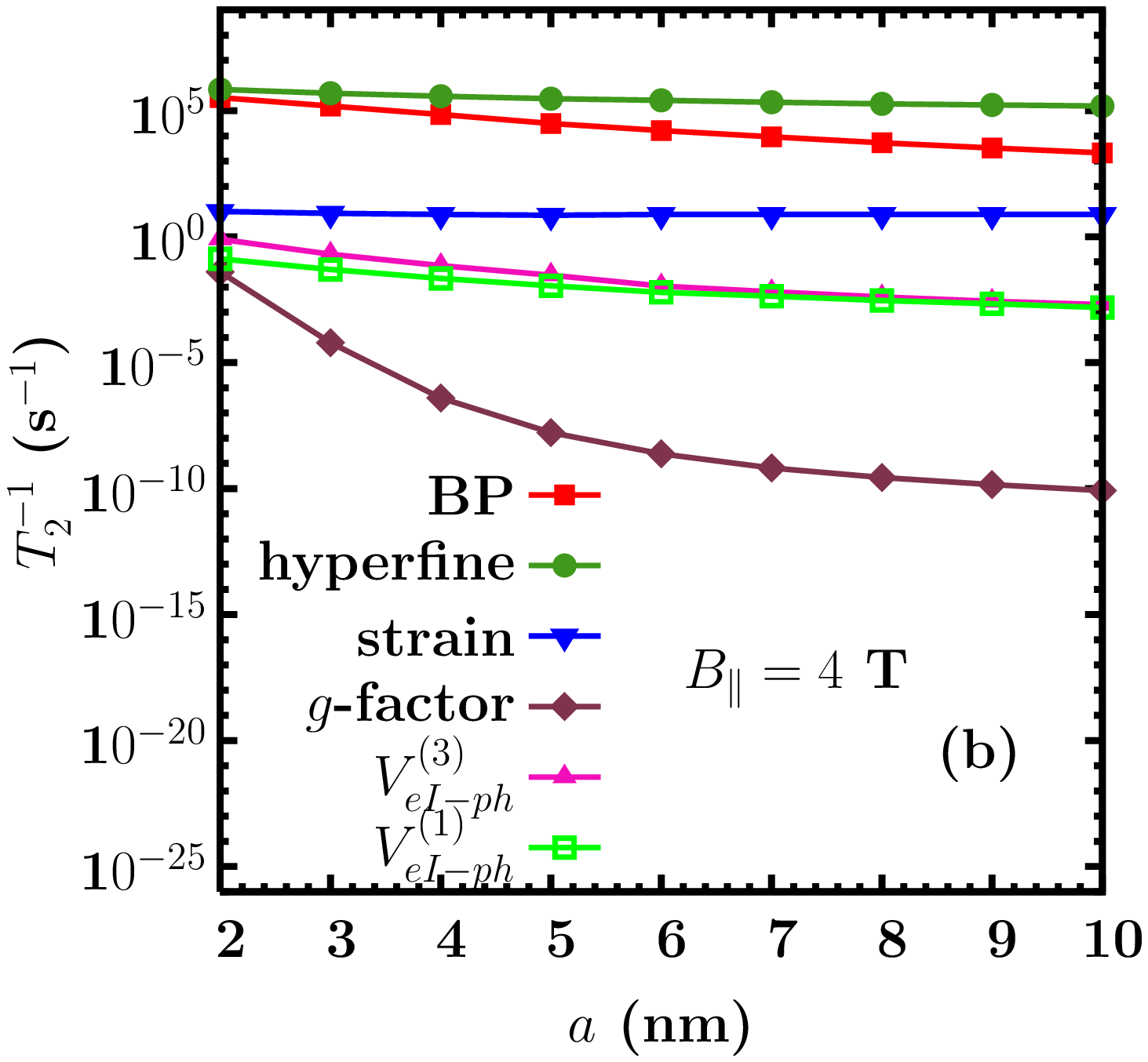}
\end{center}
\caption{(Color online) $T_{2}^{-1}$ induced by different mechanisms
  {\em vs}. the well width for $d_{0}=20$\ 
nm. $T=4$\ K. (a): $B_{\perp}=4$\ T\ with (solid curves) and without
(dashed curves) the SOC; (b): $B_{\|}=4$\ T only with the SOC.
Curve with $\blacksquare$ --- $T_{2}^{-1}$ induced by the electron-BP
interaction;
Curves with $\bullet$ --- $T_{2}^{-1}$ induced by the
hyperfine interaction;
Curves with $\blacktriangledown$ --- $T_{2}^{-1}$ induced by the
direct spin-phonon coupling due to phonon-induced strain; Curves with
$\blacklozenge$ --- $T_{2}^{-1}$ induced by $g$-factor
fluctuation; $\blacktriangle$ --- $T_{2}^{-1}$ induced by the
  second-order process of the
  hyperfine interaction together with the BP ($V_{eI-ph}^{(3)}$);
Curves with  $\square$ --- $T_{2}^{-1}$
 induced by the first-order process of the hyperfine interaction
together with the BP ($V_{eI-ph}^{(1)}$). The time evolution of
$||\langle S_{+}\rangle_t||$ induced by the hyperfine 
interaction with $a=2$\ nm is shown in the inset of (a).}
\label{fig7}
\end{figure}

In Fig.\ \ref{fig7} the well width
dependence of the spin dephasing induced by different mechanisms is
presented under the perpendicular (a) and parallel (b) magnetic fields.
In the calculations $B_{\perp}=4$\ T/$B_{\|}=4$\ T and $d_{0}=20$\ nm. 
It can be seen in both figures that the spin dephasing due to each
mechanism decreases with $a$. Moreover, the spin dephasing due to the
electron-BP scattering decreases much faster than that due to the
hyperfine interaction. These features can be understood as
following. The spin dephasing due to electron-BP scattering depends
crucially on the SOC. As the SOC is proportional to $a^{-2}$, the spin
dephasing decreases fast with $a$. For the hyperfine interaction, from
Eq.\ (\ref{t0d}) one can deduce that the decay rate of 
$||\langle S_{+}\rangle_t||$ is mainly determined by the factor
  $1/(a_{z}d_{\|}^{2})$ (here $a_{z}=a$), which thus decreases with  
$a$, but in a very mild way. The fast decrease of the electron-BP
mechanism makes it eventually unimportant. For the present
perpendicular-magnetic-field case the threshold is around 2\ nm. For
parallel magnetic field it is even smaller. A higher temperature may
enhance the electron-BP mechanism (see discussion in Sec.\ V)
and make it more important than the hyperfine mechanism.
It is noted that other mechanisms contribute very little to
the spin dephasing. Thus, in the following discussion, we do not consider
these mechanisms.  Comparing Figs\ \ref{fig7}(a) and (b),
one finds that a main difference 
is that the electron-BP mechanism is less effective for
the  parallel-magnetic-field case. 
As has been discussed in the previous subsection, the spin
mixing and the Zeeman splitting in the parallel filed case is smaller than 
those in the perpendicular field case. Therefore, the electron-BP
mechanism is weakened markedly. 

\begin{figure}[bth]
 \begin{center}
\includegraphics[height=5.5cm]{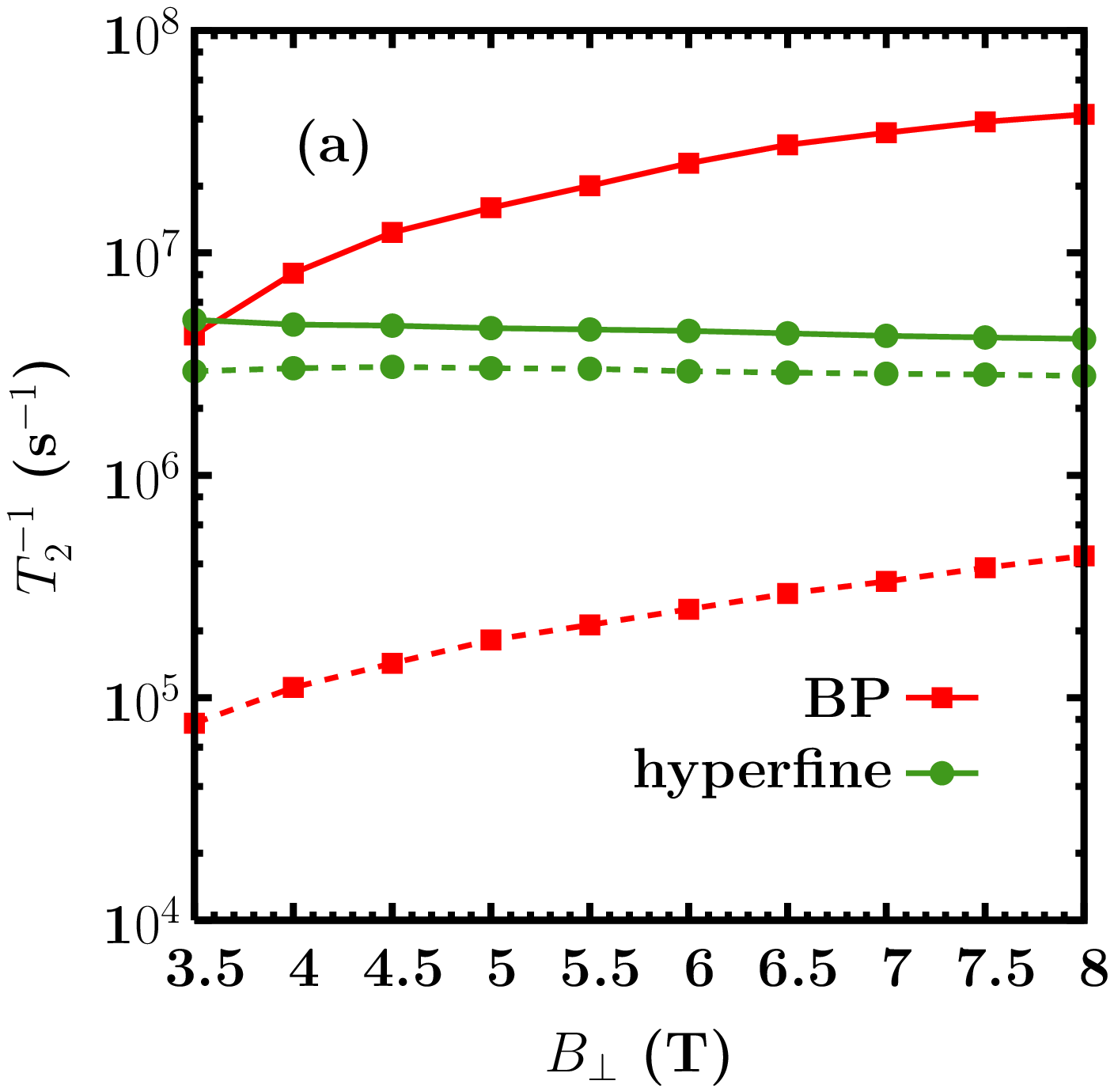}
\includegraphics[height=5.5cm]{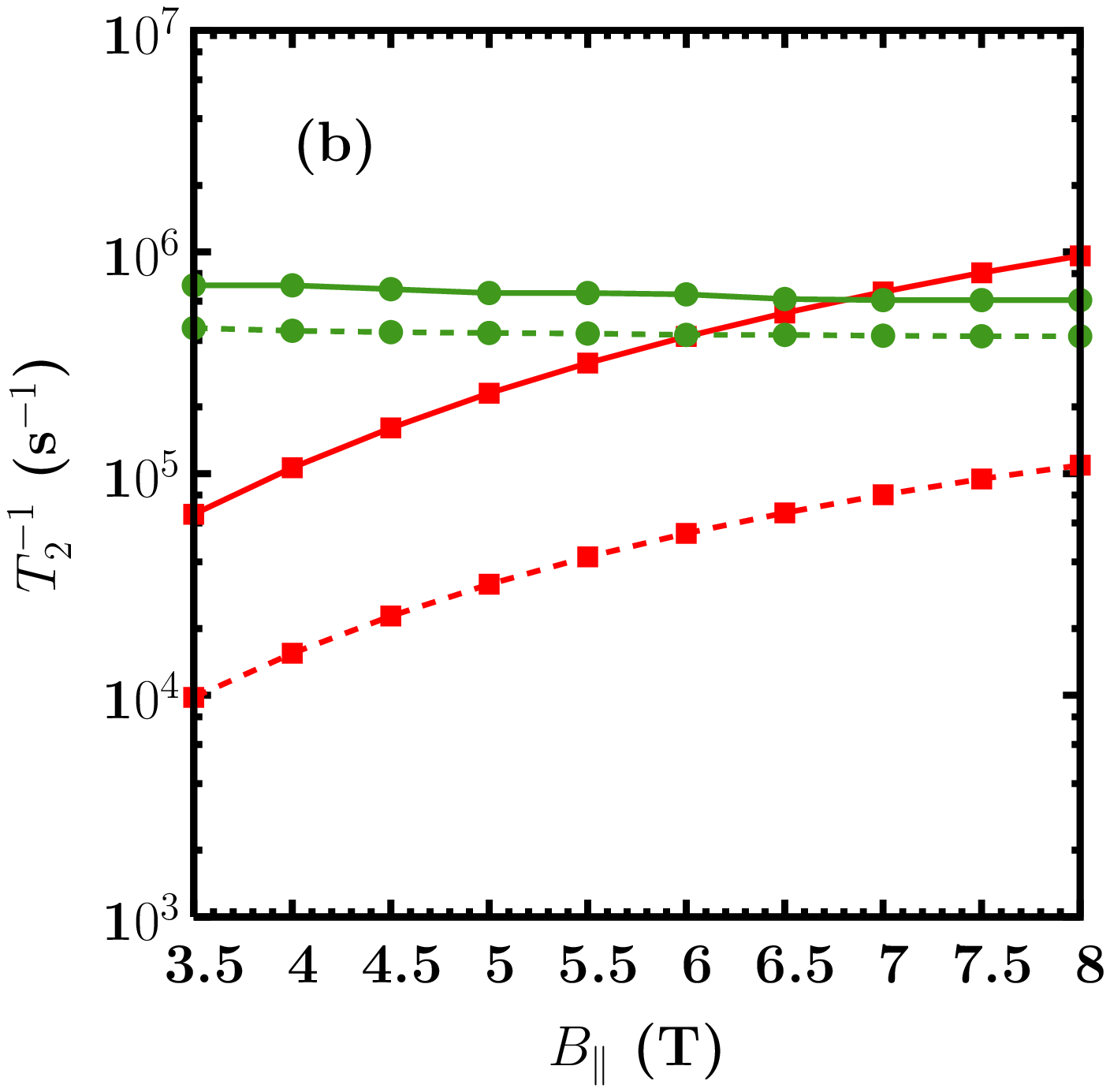}
\end{center}
\caption{(Color online) $T_{2}^{-1}$ induced by the electron-BP
  scattering and the hyperfine interaction
{\em vs}. (a): the perpendicular magnetic field $B_{\perp}$\ ; (b): 
the parallel magnetic field $B_{\|}$ for $a=3$\ nm
(solid curves) and $5$\ nm (dashed curves). $T=4$\ K, and
  $d_0=20$\ nm.
 Curves with  $\blacksquare$ --- $T_{2}^{-1}$ induced by the electron-BP
 interaction;
  Curves with $\bullet$ --- $T_{2}^{-1}$ induced by the hyperfine interaction.}
\label{fig8}
\end{figure}

Similar to Fig.\ \ref{fig1}, the SOC is always included in the
computation as it has large effect on  the eigen-energy and
eigen-wavefunction of the electrons. The spin dephasings
calculated without the SOC for the hyperfine
interaction, the direct spin-phonon coupling due to phonon-induced
strain and the $g$-factor fluctuation are also shown in Fig.\ \ref{fig7}(a) 
as dashed curves. It can be seen from the figure that for
the spin dephasings induced by the direct spin-phonon coupling due to
phonon-induced strain and by the $g$-factor fluctuation, the contributions
with the SOC are much larger than those without.
This is because when the SOC is included, the fluctuation of the
effective field induced by both mechanisms becomes much stronger and
more scattering channels are opened.
However, what should be emphasized is that the spin
dephasings induced by the hyperfine interaction
with and without the SOC are nearly the
same (the solid and the dashed curves nearly coincide). That is
because the change of the wavefunction $\Psi({\bf r})$
due to the SOC is very small (less
than 1\ \% in our condition) and therefore the factor $1/(a_{z}d_{\|}^{2})$
is almost unchanged when the SOC is neglected. Thus the spin dephasing
rate is almost unchanged.

In the inset of Fig.\ \ref{fig7}(a), 
the time evolution of $||\langle S_{+}\rangle_t||$
induced by the hyperfine interaction is shown, with  $a=2$\ nm.
It can be seen that $||\langle
  S_{+}\rangle_t||$  decays very fast and decreases to
less than 10 \% of its initial value within the first two oscillating
periods. Therefore, $T_{2}$ is determined by
the first two or three periods of $||\langle S_{+}\rangle_t||$. Thus the
correction of the long time dynamics due to higher order
scattering\cite{Coish} contributes little to the spin dephasing
time. For quantum computation and quantum information processing, 
the initial, {\sl e.g.},  1\ \% decay of 
$||\langle S_{+}\rangle_t||$  may be
more important than the $1/e$ decay.\cite{Yao,Witzel2} Indeed, the spin dephasing time
defined by the exponential fitting of  1\ \% decay is short than that
defined by the $1/e$ decay. However, the two differs less than 5
times. For a rough comparison of contributions from different mechanisms
to spin dephasing where only the order-of-magnitude difference is concerned (see
Figs.\ 7-9), this difference due to the
definition does not jeopardize our conclusions.

\subsubsection{Magnetic Field Dependence}

We then investigate the magnetic field dependence of the
spin dephasing induced by the electron-BP scattering and by the
hyperfine interaction for two different well widths ($a=3$\ nm and
$a=5$\ nm) with both perpendicular and parallel magnetic field. From
Fig.\ \ref{fig8}(a) and (b) one can see that the spin 
dephasing due to the electron-BP scattering increases with magnetic
field, whereas that due to the hyperfine interaction decreases with
magnetic field. Thus, the electron-BP mechanism  eventually dominates
the spin dephasing for high enough magnetic field. The threshold is 
$B_{\perp}^c=4$\ T / $B_{\|}^c=7$\ T for $a=3$\ nm with
perpendicular/parallel magnetic field. For larger well width,
{\em e.g.}, $a=5$\ nm with parallel magnetic field or 
 perpendicular magnetic field, the threshold magnetic fields increase
 to larger than $8$\ T. The different magnetic field dependences 
above can be understood as 
following. Besides spin relaxation, the spin-flip
scattering also contributes to spin dephasing.\cite{Golovach} As has
been demonstrated in Sec.\ IIIA, the electron-BP scattering induced
spin-flip transition rate increases 
with the magnetic field. Therefore the spin dephasing rate
increases with the magnetic field also. In contrast, spin dephasing
induced by the hyperfine interaction decreases with the magnetic field.
This is because when the magnetic field becomes larger, 
the fluctuation of the effective magnetic field
due to the surrounding nuclei becomes insignificant compared.
Therefore, the hyperfine-interaction-induced spin dephasing is reduced. Similar
results have been obtained by Deng and Hu.\cite{Deng2}

\begin{figure}[bth]
\begin{center}
\includegraphics[height=5.5cm]{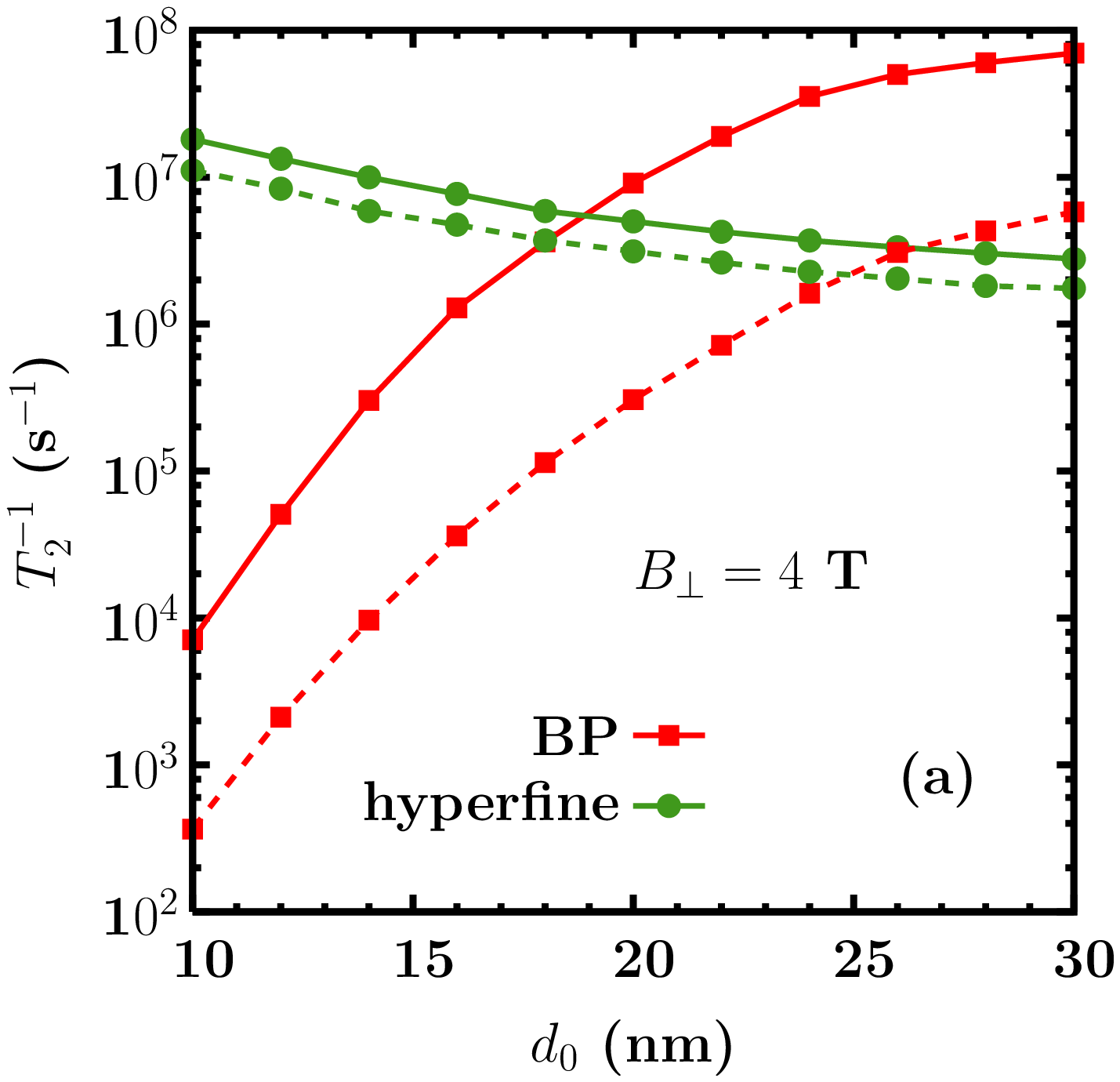}
\includegraphics[height=5.5cm]{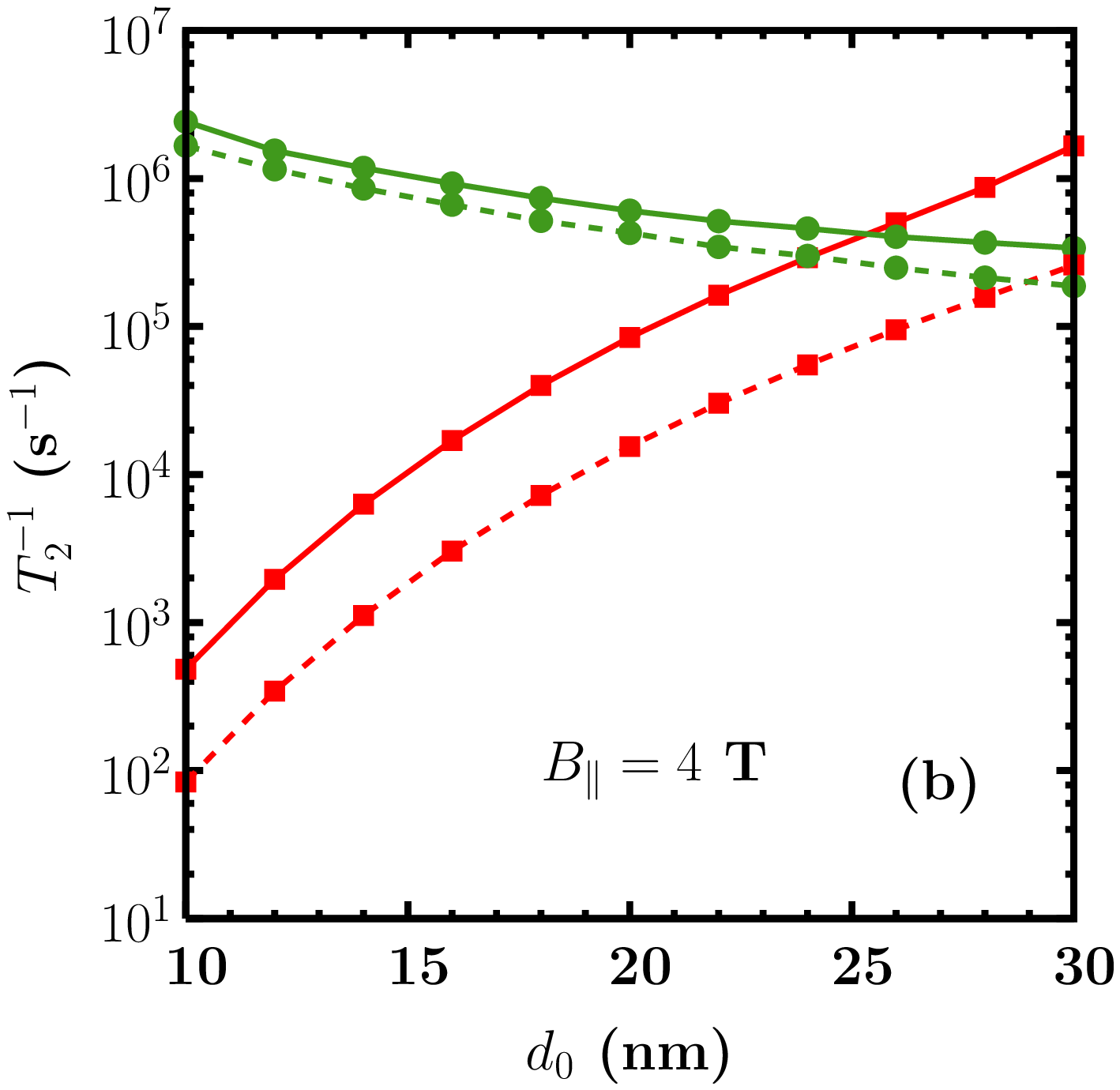}
\end{center}
\caption{(Color online) $T_{2}^{-1}$ induced by the electron-BP
  scattering and the hyperfine interaction
  {\em vs}. the effective diameter $d_{0}$ $T=4$\ K. (a): $B_{\perp}=4$\
  T\ ; (b): $B_{\|}=4$\ T for $a=3$\ nm (solid curves) and $5$\ nm
  (dashed curves). Curves with $\blacksquare$ --- $T_{2}^{-1}$  
  induced by the electron-BP interaction;
  Curves with $\bullet$ --- $T_{2}^{-1}$ induced by the hyperfine
  interaction.}
\label{fig9}
\end{figure}

\subsubsection{Diameter Dependence}
In Fig.\ \ref{fig9} the spin dephasing times induced by the
electron-BP scattering and the hyperfine interaction are plotted as
function of the diameter $d_{0}$ for a small ($a=3$\ nm) and a large
($a=5$\ nm) well widths. In the calculation, $B_{\perp}=4$\ T in  
Fig.\ \ref{fig9}(a) and $B_{\|}=4$\ T in (b). 
It is noted that the effect of the
electron-BP mechanism increases rapidly with $d_0$, whereas the effect
of the hyperfine mechanism decreases slowly. Consequently,
the electron-BP mechanism
eventually dominates the spin dephasing for large enough
$d_0$. The threshold is $d_0^c=19$ (27)\ nm for $a=3$ (5)\ nm case
with the perpendicular magnetic field and
$d_0^c=26$ (30)\ nm for $a=3$ (5)\ nm case
under the parallel magnetic field.
As has been discussed in Sec.\ IIIA, both the effect of the SOC 
and the efficiency of the phonon absorption/emission increase
with $d_0$. Therefore, the spin dephasing due to the electron-BP mechanism
increases rapidly with $d_0$.\cite{Cheng,Destefani} 
The decrease of the effect of the hyperfine 
interaction is due to the decrease of the factor $1/(a_{z}d_{\|}^{2})$
[Eq.\ \ref{t0d}] with the diameter $d_0$. 

\section{spin relaxation times from Fermi Golden rule and from Equation of motion}

In this section, we will try to find a proper method to average
 over the transition rates from the Fermi Golden
rule, $\tau_{i\to f}^{-1}$, to give the spin
 relaxation time $T_1$. In the limit of small SOC, we rederive 
Eq.\ (\ref{goldenT}) from the equation
of motion. We further show that Eq.\ (\ref{goldenT})
fails for large SOC where a full calculation from the equation of motion
is needed.

\begin{figure}[bth]
\begin{center}
\includegraphics[height=5.5cm]{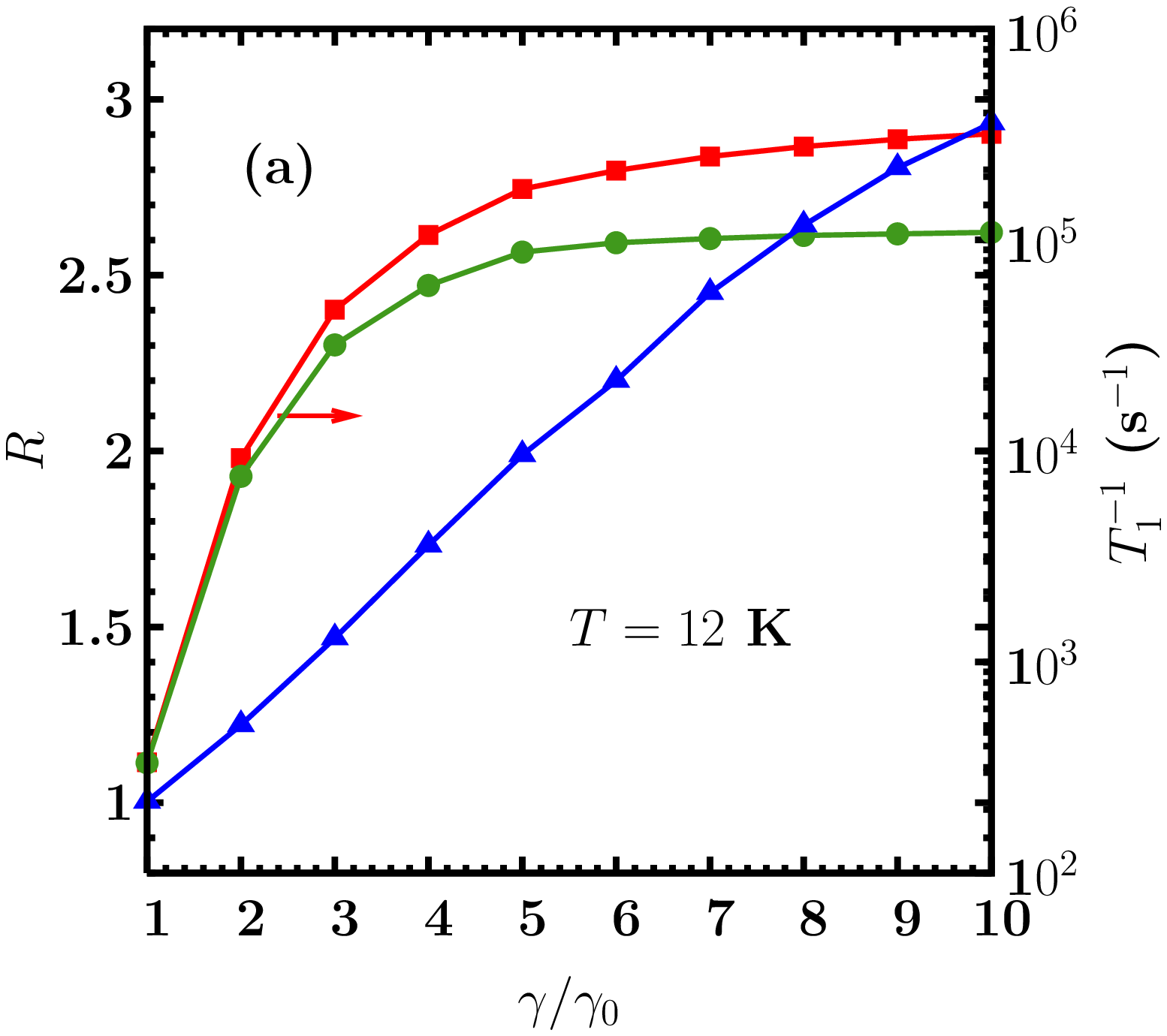}
\includegraphics[height=5.5cm]{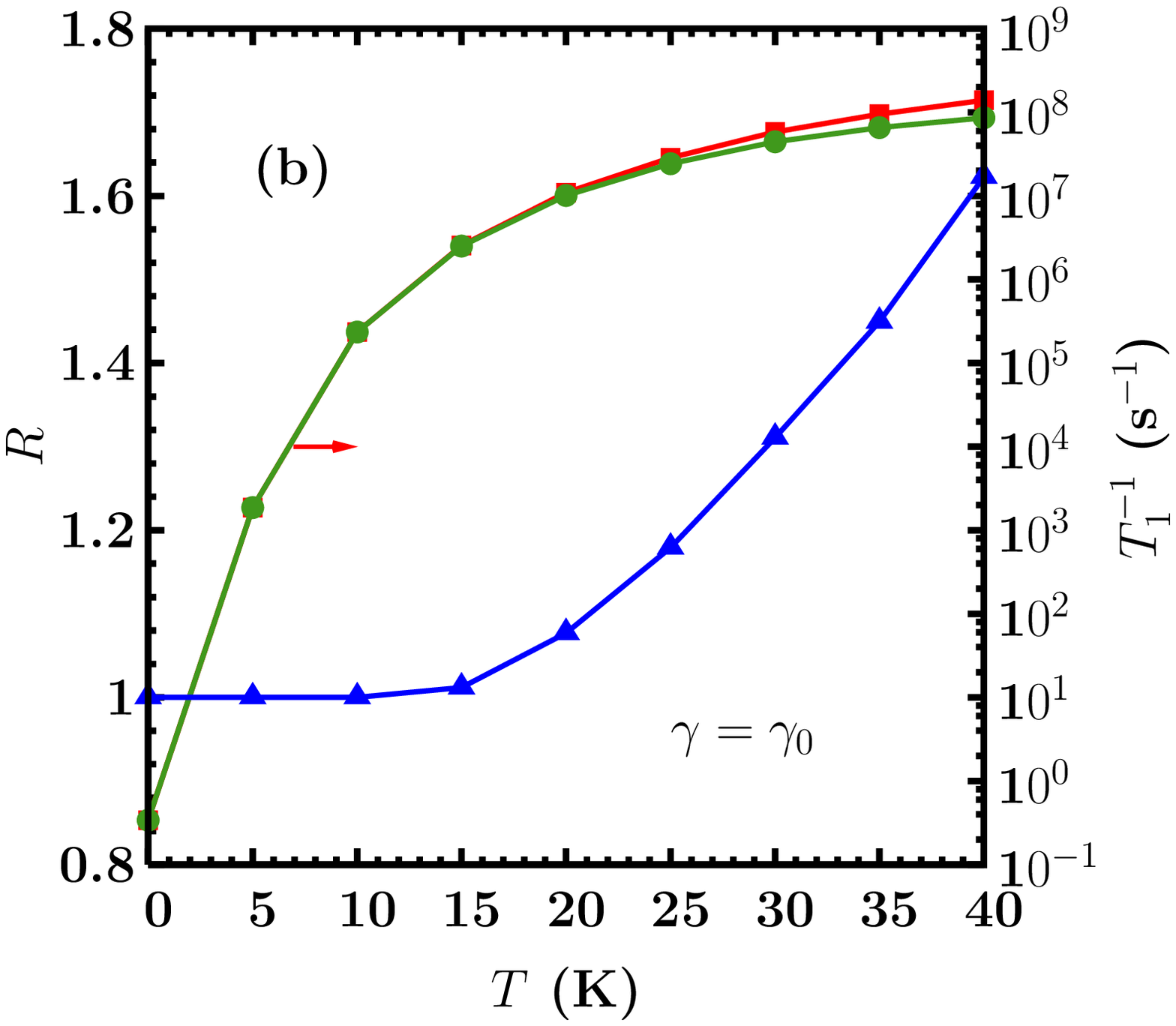}
\end{center}
\caption{(Color online) Spin relaxation time $T_1$ calculated from the
  equation-of-motion  approach ($\blacksquare$) {\em v.s.} that
  obtained from  Eq.\ (1)
  ($\bullet$) as function of (a): the strength 
  of the SOC for T=12\ K; (b): the temperature for $\gamma=\gamma_0$.
  The well width $a=5$\ nm, perpendicular magnetic field 
  $B_{\perp}=0.5$\ T, QD diameter $d_0=30$\ nm.
  The ratio of the two $R$ is also plotted in the figure.
  Note the scale of $T_1^{-1}$ is at the right hand side of the frame.}
\label{fig10}
\end{figure}

We first rederive Eq.\ (\ref{goldenT}) for small SOC
 from the equation of motion.
In QDs, the orbital level splitting is usually much larger than the
Zeeman splitting. Each Zeeman sublevel has two states: one with majority
up-spin, the other with majority down-spin. We call the former ``minus
state'' (as it corresponds to a lower energy) while the latter ``plus
state''. For small SOC, the spin mixing is small. Thus we neglect the
much smaller contribution from the off-diagonal terms of the density
matrix to $S_z$. Therefore $S_z(t) = \sum_{i\pm} S_z^{i\pm}
f_{i\pm}(t)$ where $i\pm$ denotes the 
plus/minus state of the $i$-th orbital state. For small SOC, the spin
relaxation is much slower than the orbital
relaxation.\cite{Fujisawa,Stano2} This implies that the time takes 
to establish equilibrium within the plus/minus states is much
smaller than the spin relaxation time. Thus we can assume a
equilibrium (Maxwell-Boltzmann) distribution between the plus/minus
states at any time. The distribution function is therefore given by 
$f_{i\pm}(t) = N_{\pm}(t)\exp(-\varepsilon_{i\pm}/k_BT)/Z_{\pm}$. Here
$N_{\pm}(t)=\sum_{i}f_{i\pm}(t)$ is the total probability of the
plus/minus states with $N_{+}(t)+N_{-}(t)=1$ for single electron in QD and
$Z_{\pm}=\sum_{i}\exp(-\varepsilon_{i\pm}/k_BT)$ is the 
partition function for the plus/minus state. At equilibrium,
$N_{\pm}=N_{\pm}^{eq}$. The equation for $S_z(t)$ is hence,
\begin{eqnarray}
 \hspace{-0.6cm}\frac{d}{dt} S_z(t) &=& \frac{d}{dt} [S_z(t)-S_z^{eq}]\nonumber\\
 \hspace{-0.6cm}&=& \sum_{i\pm} S_z^{i\pm} \exp(-\varepsilon_{i\pm}/k_BT)/Z_{\pm}
  \frac{d}{dt} \delta N_{\pm}(t)\ ,
  \label{mfgr1}
\end{eqnarray}
with $\delta N_{\pm}(t)=N_{\pm}(t)-N_{\pm}^{eq}$.
As the orbital level splitting is usually much larger than the Zeeman
splitting, the factor $\exp(-\varepsilon_{i\pm}/k_BT)/Z_{\pm}$ can
be approximated by $\exp(-\varepsilon_{i0}/k_BT)/Z_{0}$ with
$\varepsilon_{i0}=\frac{1}{2}(\varepsilon_{i+}+\varepsilon_{i-})$
and $Z_{0}=\sum_{i}\exp[-\varepsilon_{i0}/k_BT]$. Further using the 
particle-conservation relation
$\sum_{\pm}\delta N_{\pm}(t)=0$, one has
\begin{equation}
  \frac{d}{dt} S_z(t) = [\sum_{i} (S_z^{i+} -S_z^{i-})
\exp(-\varepsilon_{i0}/k_BT)/Z_{0}]
  \frac{d}{dt} \delta N_{+}(t)\ .
\end{equation}
As $S_z(t) - S_z^{eq}=  [\delta N_{+}(t)/Z_0]\sum_{i} (S_z^{i+}
-S_z^{i-})\exp(-\varepsilon_{i0}/k_BT)$, 
one finds that the spin relaxation time is nothing but the relaxation time of
$N_{+}$. The next step is to derive the equation of $\frac{d}{dt}
\delta N_{+}(t)$, which is given in our previous work:\cite{JHJiang}
\begin{eqnarray}
&&\frac{d}{dt} \delta N_{+}(t) = \sum_{i} \frac{d}{dt} \delta
  f_{i+}(t)\nonumber \\
&&= - \sum_{i,f}[
  \tau_{{i+}\to{f-}}^{-1} \delta f_{i+}(t) - 
  \tau_{{i-}\to{f+}}^{-1} \delta f_{i-}(t) ] \nonumber\\
&&= - \sum_{i,f} [ \tau_{{i+}\to{f-}}^{-1}
  + \tau_{{i-}\to{f+}}^{-1} ]
  \frac{e^{-\varepsilon_{i0}/k_BT}}{Z_{0}} 
\delta N_{+}(t)\ .
\end{eqnarray}
Thus spin relaxation time is given by,
\begin{equation}
 \frac{1}{T_1} = \sum_{i,f} (\tau_{{i+}\to{f-}}^{-1}
  + \tau_{{i-}\to{f+}}^{-1} )
  \frac{e^{-\varepsilon_{i0}/k_BT}}{Z_{0}}\ .
\end{equation}
Furthermore, substituting
$e^{-\varepsilon_{i0}/k_BT}/Z_{0}$ by 
$f_{i\pm}^{0}=\exp(-\varepsilon_{i\pm}/k_BT)/Z_{\pm}$, we have
\begin{equation}
 \frac{1}{T_1} = \sum_{i,f}( \tau_{{i+}\to{f-}}^{-1} f_{i+}^{0}
  + \tau_{{i-}\to{f+}}^{-1} f_{i-}^{0})\ .
\label{tau1}
\end{equation}
This is exactly Eq.\ (\ref{goldenT}).

\begin{figure}[bth]
\begin{center}
\includegraphics[height=5.5cm]{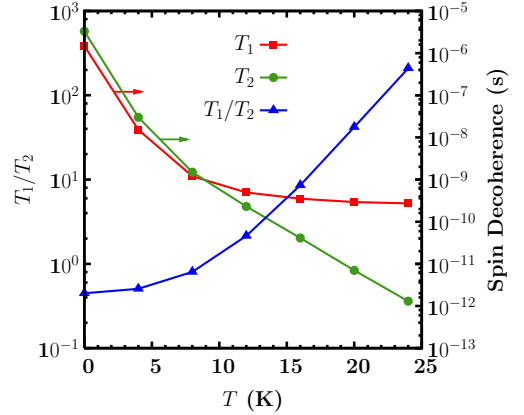}
\end{center}
\caption{(Color online) Spin relaxation time $T_1$, spin
dephasing time $T_2$ and $T_1/T_2$ against temperature $T$.
$B_{\perp}=4$\ T, $a=5$\ nm and $d_{0}=30$\ nm.
Note the scale of $T_1$ and $T_2$ is at the right hand side of the
frame.}
\label{fig11}
\end{figure}

For large SOC, or large spin mixing due to anticrossing of different
spin states,\cite{Bulaev,Stano2} the spin relaxation rate becomes
comparable with the orbital relaxation rate. Furthermore, the decay of
the off-diagonal term of the density matrix should contribute to the
decay of $S_z$. Therefore, the above analysis does not hold. In this
case, it is difficult to obtain such a formula, and a full calculation
from the equation-of-motion is needed.

In Fig.\ \ref{fig10}(a), we show that 
(for $T=12$\ K, $a=5$\ nm, $B_{\perp}=0.5$\
T, $d_0=30$\ nm) the spin relaxation times $T_1$ calculated from
equation-of-motion approach and that obtained from Eq.\ 
(\ref{tau1}). Here, for simplicity and without loss of generality, we
consider only the electron-BP scattering mechanism. The discrepancy of
$T_1$ obtained from the two approaches increases with $\gamma$. At
$\gamma=10\gamma_0$, the ratio of the two becomes as large as $\sim
3$. In Fig.\ \ref{fig10}(b), we plot the spin relaxation times obtained
via the two approaches as function of temperature for $\gamma=\gamma_0$
with other parameters remaining unchanged. It is noted that the
discrepancy of $T_1$ obtained from the two approaches increases with
temperature. For high temperature, the higher levels are involved in
the spin dynamics where the SOC becomes larger. 
At 40\ K, the discrepancy is as large as
60\ \%. The ratio increases very 
slowly for $T<20$\ K where only the lowest two Zeeman sublevels are
involved in the dynamics.

\begin{figure}[bth]
\begin{center}
\includegraphics[height=5.5cm]{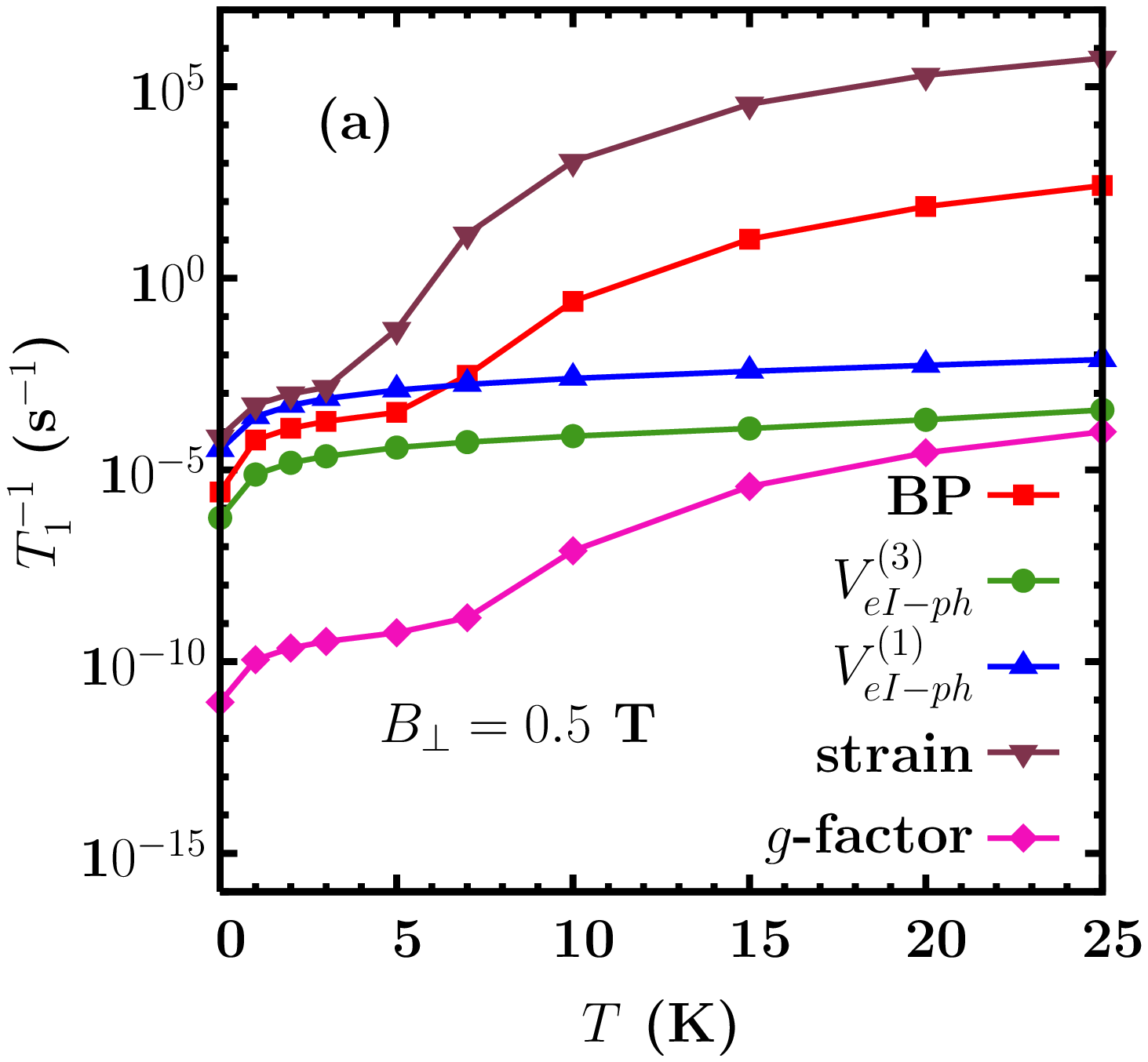}
\includegraphics[height=5.5cm]{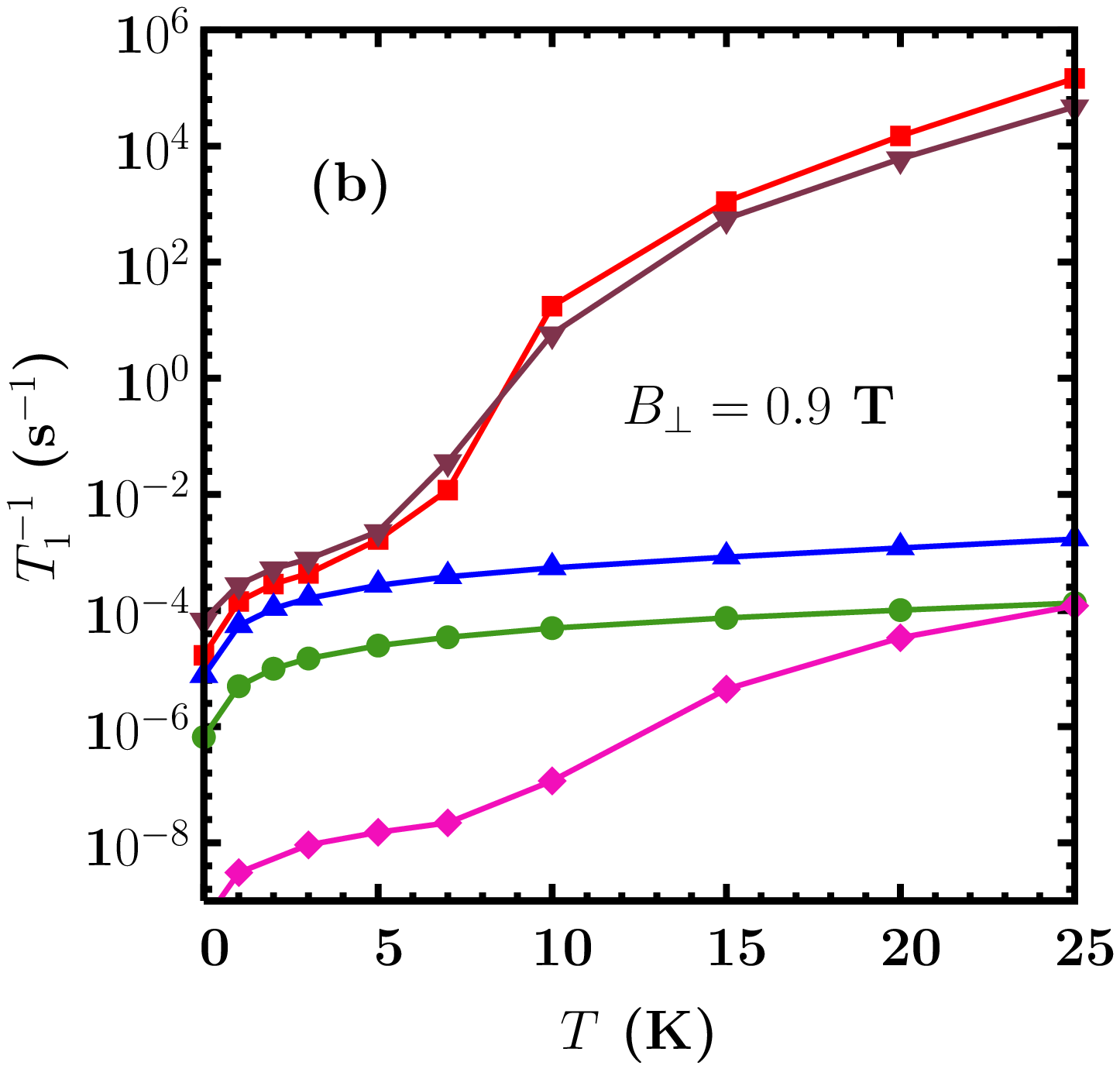}
\end{center}
\caption{(Color online) Spin relaxation time $T_1$
against temperature $T$ for (a): $B_{\perp}=0.5$\ T; (b):
$B_{\perp}=0.9$\ T. $a=10$\ nm and 
$d_{0}=20$\ nm. Curves with $\blacksquare$ --- $T_{1}^{-1}$ induced by the electron-BP
scattering together with the SOC;
Curves with $\bullet$ --- $T_{1}^{-1}$ induced by the
  second-order process of the
  hyperfine interaction together with the BP ($V_{eI-ph}^{(3)}$);
Curves with
  $\blacktriangle$ --- $T_{1}^{-1}$
 induced by the first-order process of the hyperfine interaction
together with the BP ($V_{eI-ph}^{(1)}$);
Curves with $\blacktriangledown$ --- $T_{1}^{-1}$ induced by the
 direct spin-phonon coupling due to phonon-induced strain; Curves
with $\blacklozenge$ --- $T_{1}^{-1}$ induced by the $g$-factor fluctuation.}
\label{fig12}
\end{figure}

\section{Temperature Dependence of Spin
relaxation time $T_{1}$ and  spin dephasing time $T_{2}$}

We first study the relative magnitude of the
spin relaxation time $T_{1}$ and the spin dephasing time $T_{2}$. We
consider a QD with  $d_{0}=30$\ nm and $a=5$\ nm at $B_\perp=4$\ T
 where the largest
contribution to both  spin relaxation and  dephasing comes from the
electron-BP scattering (see Fig.\ \ref{fig4}(a) and Fig.\
\ref{fig9}(a), we have checked that the electron-BP scattering
  mechanism is dominant throughout the temperature range). 
From Fig.\ \ref{fig11}, one finds that
when the temperature is low ($T<5$\ K in the figure),
$T_{2}=2T_1$, which is in agreement with the discussion in
Ref.\ \onlinecite{Golovach}.  However, $T_{1}/T_{2}$ increases very
quickly with $T$ and for $T=20$\ K, $T_{1}/T_{2} \sim 2 \times
10^{2}$. This is understood from the fact that
when $T$ is low, the electron mostly distributes in the lowest two
Zeeman sublevels. 
For small SOC, Golovach {\em et al.} have shown via perturbation
theory that phonon induces only the spin-flip noise in the leading
order. Consequently, $T_2=2T_1$.\cite{Golovach}
When the temperature becomes comparable with the orbital level
splitting $\hbar\omega_0$, the distribution over the upper
orbital levels is not negligible any more. As mentioned previously,
the SOC contributes a non-trivial part to the Zeeman
splitting. Specifically, the second order energy correction due to the
SOC contributes to the Zeeman splitting. The energy correction for
different orbital levels is generally unequal (always larger for
  higher levels).
When the electron is scattered by phonons randomly from one
orbital state to another one with the same major spin polarization,
the frequency of its precession around $z$ direction
changes. Continuous scattering leads to random fluctuation of the
precession frequency and thus leads to spin
dephasing.\cite{Semenov,Semenov3}
Note that this fluctuation only leads to a phase randomization of
$S_{+}$, but not flips the $z$ component spin $S_z$, {\em i.e.}, not
leads to spin relaxation.
Therefore, the spin dephasing becomes stronger than the
spin relaxation for high temperatures. Moreover, this effect
  increases with temperature rapidly as the distribution 
over higher levels and the phonon numbers both increase with temperature.

We further study the temperature dependence of spin relaxation for
  lower magnetic field and larger quantum well width where other
  mechanisms may be more important than the electron-BP mechanism. 
In Fig.\ \ref{fig12}(a), the spin relaxation time is plotted as
function of temperature for $B_{\perp}=0.5$\ T, $a=10$\ nm and
$d_{0}=20$\ nm. It is seen from the figure that the direct spin-phonon
coupling due to phonon-induced strain mechanism dominates the spin
relaxation throughout the temperature range. It is also noted that for
$T\le 4$\ K the spin relaxation rates induced by different mechanisms
all increase with temperature according to the phonon number factor
$2\bar{n}(E_{z1})+1$ with $E_{z1}$ being the Zeeman splitting of the lowest
Zeeman sublevels. However, for $T> 4$\ K, the spin relaxation rates
induced by the direct spin-phonon coupling due to phonon-induced
strain and the electron-BP interaction increase rapidly with
temperature, while the spin relaxation rates induced by 
$V_{eI-ph}^{(1)}$ and $V_{eI-ph}^{(3)}$ increase mildly 
according to $2\bar{n}(E_{z1})+1$ 
throughout the temperature range. 
These features can be understood as what follows.
 For $T\le 4$\ K, the distribution over the high levels is
negligible. Only the lowest two Zeeman sublevels involve in the spin
dynamics. The spin relaxation rates thus increase with
$2\bar{n}(E_{z1})+1$ and the relative importance of each mechanism
does not change. Therefore, our previous analysis on comparison of
relative importance of different spin decoherence mechanisms at 4\ K
holds true for the range $0\le T\le 4$\ K.  When the temperature gets
higher, the contribution from  higher levels becomes more 
important. Although the distribution at the higher levels is still very
small, for the direct spin-phonon coupling mechanism, the transition
rates between the higher levels and that between higher levels and the
lowest two sublevels are very large. For the electron-BP mechanism the
transition rates between the higher levels are very large 
due to the large  SOC 
 in these  levels. Therefore, the 
contribution from the higher levels becomes larger than that from the
lowest two sublevels. Consequently, the increase of temperature leads
to rapid increase of the spin relaxation rates. However,
for the two hyperfine mechanisms: the $V_{eI-ph}^{(1)}$ and the
$V_{eI-ph}^{(3)}$, the spin relaxation rates does not change much 
when the higher levels are involved. They thus  increase
 by the phonon number factor. 

In Fig.\ \ref{fig12}(b) we show the temperature dependence of the spin
relaxation time for the same condition but with $B=0.9$\ T. It is
noted that the spin relaxation rate due to the electron-BP mechanism
catches up with that induced by the direct spin-phonon coupling due to
phonon-induced strain at $T=9$\ K and becomes larger for higher
temperature. This indicates that the temperature dependence of
  the two mechanisms are quite different.

\begin{figure}[bth]
\begin{center}
\includegraphics[height=5.5cm]{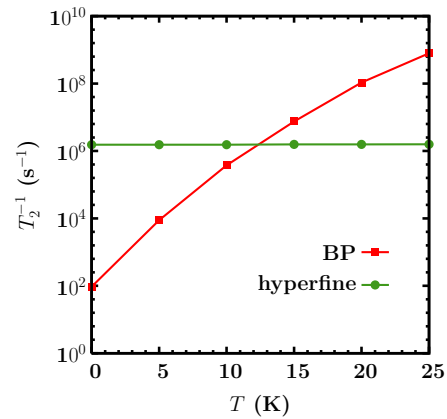}
\end{center}
\caption{(Color online) Spin relaxation time $T_1$
against temperature $T$. $B_{\perp}=4$\ T, $a=10$\ nm and
$d_{0}=20$\ nm. Curves with $\blacksquare$ --- $T_{2}^{-1}$ induced by
the electron-BP scattering together with the SOC;
Curves with $\bullet$ --- $T_{2}^{-1}$ induced by the
hyperfine interaction}
\label{fig13}
\end{figure}

In Fig.\ \ref{fig13} we show the spin dephasing  induced by
 electron-BP scattering and the hyperfine interaction 
as function of temperature for  $B_{\perp}=4$\ T, $a=10$\ nm and
$d_{0}=20$\ nm. We choose the conditions so that the spin dephasing
is dominated by the hyperfine interaction at low
temperature. However, the effect of the electron-BP mechanism
increases with temperature quickly while that of the hyperfine
interaction remains nearly unchanged. The fast increase of the effect 
from the  electron-BP scattering is due to three factors: 
1) the increase of the
phonon number; 2) the increase of scattering channels; and
3) the increase of the SOC induced spin mixing in higher levels.
On the other hand, from Eq.\ \ref{t0d}, one can 
deduce that the spin dephasing rate of the hyperfine interaction
depends mainly on the factor $1/(a_{z} d_{\|}^2)$ with
$a_{z}$/$d_{\|}^2$ is the characteristic length/area along the $z$
direction / in the quantum well plane. For higher levels, the
$d_{\|}^2$ is larger, but only about a factor smaller than 10. 
Thus the effect of the hyperfine interaction
increases very slowly with temperature.

It should be noted that in the above discussion,
we neglected the two-phonon scattering
 mechanism,\cite{Glavin,Semenov3,Nazarov} 
which may be important at high temperature.
The contribution of this mechanism should be calculated via the
equation-of-motion approach developed in this paper, and compared with
the contribution of other mechanisms showed here.

\section{Conclusion}

In conclusion, we have investigated the longitudinal and transversal
spin decoherence times $T_{1}$ and $T_{2}$, called spin
relaxation time and spin dephasing time, in different conditions
in GaAs QDs from the equation-of-motion approach. Various
mechanisms, including the electron-BP scattering, the
hyperfine interaction, the direct spin-phonon coupling due to phonon-induced
strain and the $g$-factor fluctuation are considered. 
Their relative importance is compared. 
There is no doubt that for spin decoherence induced 
by electron-BP scattering, the SOC must be included.
However, for spin decoherence induced by the hyperfine interaction, 
the direct spin-phonon coupling due to phonon-induced strain,
$g$-factor fluctuation, and hyperfine interaction combined with
electron-phonon scattering, the SOC is neglected in the existing
literature.\cite{Erlingsson,Abalmassov,Kim} 
Our calculations have shown that, as the SOC has marked effect on the
eigen-energy and the eigen-wavefunction of the electron, the spin
decoherence induced by these mechanisms with the SOC is larger than that
without it. Especially, the decoherence from
the  second-order process of hyperfine interaction
combined with the electron-BP interaction increases at least
one order of magnitude when the SOC is included. 
Our calculations show that, with the
SOC, in some conditions some of these mechanisms (except $g$-factor
fluctuation mechanism) can even dominate the spin decoherence.

There is no single mechanism which dominates spin relaxation
or spin dephasing in all parameter regimes. 
The relative importance of each mechanism varies with the well width,
magnetic field and QD diameter. 
In particular, the electron-BP scattering mechanism has the largest
contribution to spin relaxation and spin dephasing for small 
well width and/or high  magnetic field and/or large QD
diameter. However, for other parameters the hyperfine interaction, the
first-order process of the hyperfine interaction combined with
electron-BP scattering, and the direct spin-phonon coupling due to
phonon-induced strain can be more important. 
It is noted that the $g$-factor fluctuation always has very little
contribution to spin relaxation and spin dephasing which can thus be
neglected all the time. For spin dephasing, the electron-BP 
scattering mechanism
and the hyperfine interaction mechanism are  more important than
other mechanisms for magnetic field higher than 3.5\ T. For this
regime, other mechanisms can thus be neglected. It is also shown that 
spin dephasing induced by the electron-BP mechanism increases rapidly
with temperature. Extrapolated from our calculation, the hyperfine
interaction mechanism is believed to be dominant for small magnetic
field.

We also discussed the problem of finding a proper method to average
over the transition rates $\tau_{i \to f}^{-1}$ obtained from the
Fermi Golden rule, to give the spin
relaxation time $T_1$ at finite temperature. 
For small SOC, we rederived the 
 formula for $T_1$  at finite temperature 
 used in the existing literature\cite{Cheng,lv,Wang} 
from the equation of motion.
We further demonstrated that this formula is
inadequate at high temperature and/or for large SOC. For such cases, a full
calculation from the equation-of-motion approach is needed. 
The equation-of-motion approach provides an easy and powerful way
to calculate the spin decoherence at  {\em any}  temperature and 
SOC.

We also studied the temperature dependence of spin relaxation $T_1$ and
dephasing $T_2$. We show that for very low temperature if the electron only
distributes on the lowest two Zeeman sublevels, $T_{2}=2T_1$.
However, for higher temperatures, the electron spin dephasing
increases with temperature much faster than the spin relaxation.
Consequently $T_1\gg T_2$. The spin relaxation and dephasing due to
different mechanisms are also compared.

\begin{acknowledgments}
 This work was supported by the Natural Science Foundation of China
under Grant Nos.\ 10574120 and 10725417,  the National Basic Research Program of
China under Grant No.\ 2006CB922005 and
the Innovation Project of Chinese Academy of Sciences.
Y.Y.W. would like to thank J. L. Cheng for  valuable
discussions.

\end{acknowledgments}


\begin{thebibliography}{0}
\bibitem{note1} It should be mentioned that one 
  effect is not included : when electron is scattered by phonon from
  one orbital state to another, it feels a difference in the spin
  precession frequency since the strength of longitudinal (along the
  external magnetic field) component of Overhauser field differs with 
  orbital states. This effect randomizes the spin precession phase and
  leads to a pure spin dephasing. However, this effect is negligible
  in our manuscript.
\bibitem{note2} The deviation of our calculation from the
 experiment data at $T=14$\ T is due to the fact that we do not include
  the cyclotron effect along the-$z$ direction. For $B\gtrsim 10$\ T,
 the cyclotron orbit length is smaller than the quantum well
  width, which makes our model unrealistic.
\end{thebibliography}
\end{document}